\documentclass[a4paper,12pt]{article}
\usepackage{amsfonts}
\usepackage{latexsym}
\usepackage{amssymb}
\date{}

\begin{document}

\title{Covariant Relativistic\\ Non-Equilibrium Thermodynamics of Multi-Component
Systems\thanks{In memory of Robert Trostel}
}
\author{W. Muschik\footnote{Corresponding author:
muschik@physik.tu-berlin.de}
\\
Institut f\"ur Theoretische Physik\\
Technische Universit\"at Berlin\\
Hardenbergstr. 36\\D-10623 BERLIN,  Germany}
\maketitle

            \newcommand{\be}{\begin{equation}}
            \newcommand{\beg}[1]{\begin{equation}\label{#1}}
            \newcommand{\ee}{\end{equation}\normalsize}
            \newcommand{\bee}[1]{\begin{equation}\label{#1}}
            \newcommand{\bey}{\begin{eqnarray}}
            \newcommand{\byy}[1]{\begin{eqnarray}\label{#1}}
            \newcommand{\eey}{\end{eqnarray}\normalsize}
            \newcommand{\beo}{\begin{eqnarray}\normalsize}
       
            \newcommand{\R}[1]{(\ref{#1})}
            \newcommand{\C}[1]{\cite{#1}}

            \newcommand{\mvec}[1]{\mbox{\boldmath{$#1$}}}
            \newcommand{\x}{(\!\mvec{x}, t)}
            \newcommand{\m}{\mvec{m}}
            \newcommand{\F}{{\cal F}}
            \newcommand{\n}{\mvec{n}}
            \newcommand{\argm}{(\m ,\mvec{x}, t)}
            \newcommand{\argn}{(\n ,\mvec{x}, t)}
            \newcommand{\T}[1]{\widetilde{#1}}
            \newcommand{\U}[1]{\underline{#1}}
            \newcommand{\V}[1]{\overline{#1}}
            \newcommand{\ub}[1]{\underbrace{#1}}
            \newcommand{\X}{\!\mvec{X} (\cdot)}
            \newcommand{\cd}{(\cdot)}
            \newcommand{\Q}{\mbox{\bf Q}}
            \newcommand{\p}{\partial_t}
            \newcommand{\z}{\!\mvec{z}}
            \newcommand{\bu}{\!\mvec{u}}
            \newcommand{\rr}{\!\mvec{r}}
            \newcommand{\w}{\!\mvec{w}}
            \newcommand{\g}{\!\mvec{g}}
            \newcommand{\D}{I\!\!D}
            \newcommand{\se}[1]{_{\mvec{;}#1}}
            \newcommand{\sek}[1]{_{\mvec{;}#1]}}            
            \newcommand{\seb}[1]{_{\mvec{;}#1)}}            
            \newcommand{\ko}[1]{_{\mvec{,}#1}}
            \newcommand{\ab}[1]{_{\mvec{|}#1}}
            \newcommand{\abb}[1]{_{\mvec{||}#1}}
            \newcommand{\td}{{^{\bullet}}}
            \newcommand{\eq}{{_{eq}}}
            \newcommand{\eqo}{{^{eq}}}
            \newcommand{\f}{\varphi}
            \newcommand{\rh}{\varrho}
            \newcommand{\dm}{\diamond\!}
            \newcommand{\seq}{\stackrel{_\bullet}{=}}
            \newcommand{\st}[2]{\stackrel{_#1}{#2}}
            \newcommand{\om}{\Omega}
            \newcommand{\emp}{\emptyset}
            \newcommand{\bt}{\bowtie}
            \newcommand{\btu}{\boxdot}
            \newcommand{\tup}{_\triangle}
            \newcommand{\tdo}{_\triangledown}
            \newcommand{\Ka}{\frac{\nu^A}{\Theta^A}}
            \newcommand{\K}[1]{\frac{1}{\Theta^{#1}}}
            \newcommand{\ap}{\approx}
            \newcommand{\bg}{\st{\Box}{=}}
            \newcommand{\si}{\simeq}
\newcommand{\Section}[1]{\section{\mbox{}\hspace{-.6cm}.\hspace{.4cm}#1}}
\newcommand{\Subsection}[1]{\subsection{\mbox{}\hspace{-.6cm}.\hspace{.4cm}
\em #1}}

\newcommand{\const}{\textit{const.}}
\newcommand{\vect}[1]{\underline{\ensuremath{#1}}}
\newcommand{\abl}[2]{\ensuremath{\frac{\partial #1}{\partial #2}}}


\vspace{1cm}\noindent
{\bf Abstract} Non-equilibrium and equilibrium thermodynamics of an interacting component in a relativistic multi-component system is discussed covariantly by exploiting
an entropy identity. The special case of the corresponding free component is considered.
Equilibrium conditions and especially the multi-component Killing relation of the
4-temperature are discussed. Two axioms characterize the mixture: additivity of the
energy momentum tensors and additivity of the 4-entropies of the components generating
those of the mixture. The resulting quantities of a single component and of the mixture
as a whole, energy, energy flux, momentum flux, stress tensor, entropy, entropy flux,
supply and production are derived. Finally, a general relativistic 2-component mixture is
discussed with respect to their gravitation generating energy-momentum tensors.

\section{Introduction}

The treatment of multi-component systems is often restricted to transport phenomena in
chemically reacting systems, that means, the mixture consisting of different components is
shortly
described by 1-component quantities such as temperature, pressure and energy which are not retraced to the corresponding quantities of the several components of the
multi-component system. That is the case in
non-relativistic physics \C{dGM,M85a,KBJG17} as well as in relativistic physics \C{K1,K2,IS,HASW,NEU}. In this paper, the single component as an interacting member of the mixture is investigated. Thus, each component of the mixture is equipped with its own temperature,
pressure, energy and mass density which all together generate the corresponding
quantities of the mixture.

Considering a multi-component system,  three items have to be distinguished: one
component as a member of the multi-component system which interacts with all the
other components of the system, the
same component as a free 1-component system separated from the multi-component
system and finally the multi-component system itself as a mixture which is composed of its components. Here, all three items are discussed in a covariant-relativistic framework.
For finding out the entropy-flux, -supply, -production and -density, a special tool is used:
the entropy identity which constrains the possibility of an arbitrary choice of these
quantities
\C{BOCH,RELTAG,MUBO,MUBO1}. Following J. Meixner and J.U. Keller that entropy in
non-equilibrium cannot be defined unequivocally \C{MEI,MEI1,KEL,KER,MU18}, the
entropy identity is only an (well set up) ansatz  for constructing a non-equilibrium
entropy and further corresponding quantities. This fact in mind,
a specific entropy and the corresponding Gibbs and
Gibbs-Duhem equations are derived. The definition of the rest mass flux densities, of the
energy and momentum balances and of the corresponding balances of the spin tensor
are taken into account as contraints in the entropy identity by introducing fields of
Lagrange multipliers. The physical dimensions of these factors allow to determine their
physical meaning.

Equilibrium is defined by equilibrium conditions which are divided into basic ones given
by vanishing entropy-flux, -supply and -production and into supplementary ones such as vanishing diffusion flux, vanishing heat flux and zero rest mass production
\C{MUBO,MUBO1}. The Killing
relation of the 4-temperature concerning equilibrium is shortly discussed. Constitutive
equations are out of scope of this paper.

The paper is organized as follows: After this introduction, the kinematics of a
multi-component system is considered in the next two sections for introducing the mass
flux and the diffusion flux densities. The energy-momentum tensor is decomposed
into its (3+1)-split, and the entanglement of the energy and momentum balances
are discussed, follwed by non-equilibrium thermodynamics of an interacting component
of the mixture and that of the corresponding free component. The equilibrium of
both is considered. Thermodynamics of the mixture starts with two axioms: additivity
of the energy momentum tensors and of the 4-entropies of the components resulting in
those of the mixture. Entropy, entropy flux, -supply and -production are found out. The
paper finishes discussing the gravitation generated by a special general-relativistic
2-component system: one component equipped with a symmetric energy-momentum
tensor, the other one with a skew-symmetric energy-momentum tensor. A summary
and an appendix are added.

\section{Kinematics}
\subsection{The components}

We consider a multi-component system consisting of $Z$ components.
The component index $^A$ runs from $1$ to $Z$. Each component has
its own rest frame ${\cal B}^A$ in which the rest mass density $\varrho^A$
is locally defined. These rest mass densities are relativistic invariants and therefore frame
independent\footnote{more details in Appendix \ref{RMD}}.
 
In general, the components have different 4-velocities: $u^A_k,\ A=1,2,...,Z;
k=1,...,4,$
which all are tensors of first order under Lorentz transformation. We now define the
component mass flux density as a 4-tensor of first order and the component mass production term as a scalar 
\bee{K2}
N^A_k\ :=\ \varrho^A u^A_k,\qquad N^{Ak}{_{;k}}\ =\ \Gamma^A.
\ee
Here, \R{K2}$_2$ is the mass balance equation of the $^A$-component .
Consequently, we introduce the basic fields of the components
\bee{K3}
\{\varrho^A,\ u^A_k\},\qquad A=1,2,...,Z.
\ee

The mass production term has two reasons: an external one by mass supply and one internal one
by chemical reactions
\bee{K3a} 
\Gamma^A\ =\ ^{(ex)}\Gamma^A + ^{(in)}\!\Gamma^A.
\ee
The external mass supply $^{(ex)}\Gamma^A$ depends on the environment of the system,
whereas $^{(in)}\Gamma^A$ is determined by chemical reactions depending on the
set of frame-independent stoichiometric equations which are discussed in Appendix \ref{SE}.

\subsection{The mixture\label{MIX}}

As each component, also the multi-component system has a mass density
$\varrho$ and a 4-velocity $u_k$ which are determined by the partial
quantities of the components. For deriving $\varrho$ and $u_k$, we apply
the nearly self-evident
\vspace{.3cm}\newline
$\blacksquare$\ {\sf Mixture Axiom:} The balance equation of a mixture
looks like the balance equation of an one-component system.
\hfill$\blacksquare$
\vspace{.3cm}\newline
Especially here, the mixture axiom is postulated for the balance equations of
mass, energy-momentum and entropy.
According to the mixture axiom, the mass balance of the mixture looks according to
\R{K2}$_2$  
\bee{K4}
N^k{_{;k}}\ =\ \Gamma,\qquad\Gamma\ =\ 0,
\ee
with vanishing total mass production, if the mass of the mixture is conserved\footnote{the
mixture as a closed system}. 

Now the question arises: which
quantities of the components of the mixture are additive? Obviously, neither the mass
densities $\varrho^A$ nor the 4-velocities $u^A_k$ are additive
quantities according to their definitions.
Consequently, we demand in accordance with the mixture axiom
that the mass flux densities are additive\footnote{The sign $\st{\td}{=}$ stands
for a setting and $:=$ for a definition.}
\bey\nonumber
\mbox{\sf Setting I:}\hspace{9.9cm}
\\ \label{K5}
\sum_A N^A_k\ \st{\td}{=}\ N_k\ :=\  \varrho u_k\ =\
\sum_A \varrho^A u^A_k\quad\longrightarrow\quad
u_k\ =\ \sum_A\frac{\varrho^A}{\varrho}u^A_k.
\eey
For the present, $\varrho$ and $u_k$ are unknown. Of course, they depend
on the basic fields of the components \R{K3}.
Contraction with $u^k$ and use of \R{K5}$_{2,3}$ results in
\bee{K6}
\varrho\ =\ \frac{1}{c^2}\sum_A\varrho^A u^A_ku^k\ =\ \frac{1}{c^2}N_ku^k\
=\ \frac{1}{c^2}N_k\frac{1}{\varrho}N^k\ \longrightarrow\
\varrho\ =\ \pm\frac{1}{c}\sqrt{N_kN^k},
\ee
or in more detail
\bee{K6a}
\varrho\ =\ \pm\frac{1}{c}\sqrt{\sum_{A,B}\varrho ^A\varrho^Bu^A_ku^{Bk}}.
\ee
The mass density $\varrho$ and the 4-velocity $u_k$ of the mixture are
expressed by those of the components according to \R{K6a} and \R{K5}$_4$.
According to \R{K5}$_4$, the 4-velocity of the mixture is a weighted mean value
of the 4-velocities of the components. For the mass density, we have according to
\R{K6}$_1$ also a with the Kluitenberg factor $f^A$\ \footnote{$>$0 results from the representation
of the 4-velocities in components}
weighted mean value of the mass density components \C{KLGR}
\bee{K6b}
f^A\ := \frac{1}{c^2}u^A_ku^k\ >\ 0\quad\longrightarrow\quad
\rh\ =\ \sum_Af^A\rh^A\ =\ \sum_Af^A(u^A_k,u^k)\varrho^A,
\ee
resulting in the entanglement of $\rh$ and $u_k$ which are not independent of each
other
\bee{K6c}
\rh\ =\ R(\rh^A,u^A_k,u_k),\qquad u_k\ =\ U_k(\rh^A,u^A_k,\rh).
\ee

According to \R{K5}$_1$ and \R{K2}$_2$, we obtain the additivity of the mass
production terms
\bee{K7a}
N^k{_{;k}}\ =\ \sum_AN^{Ak}{_{;k}}\ =\ \sum_A\Gamma ^A\ =\ \Gamma
\ \longrightarrow\ \sum_A {^{ex}}\Gamma ^A\ =\ ^{ex}\Gamma,\
\sum_A {^{in}}\Gamma ^A\ =\ 0.\footnote{Chemical reactions are mass conserving.}
\ee

\subsection{The diffusion flux}

From \R{K5}$_3$ and \R{K6b}$_2$ follows
\bee{M1}
0\ =\ \sum_A\rh^Au^A_k - u_k\sum_Af^A\rh^A\ =\ \sum_A \rh^A(u^A_k-f^Au_k).
\ee
Introducing the diffusion flux density using \R{M1}$_2$
\bee{K8c}
J^A_k\ :=\ \varrho^A(u^A_k - f^Au_k)\ =\ N^A_k - \rh^Af^Au_k
\quad\longrightarrow\quad
\sum_A J^A_k\ =:\ J_k\ =\ 0,
\ee
we obtain 
\byy{K9}
J^A_ku^k &=& \varrho^A(u^A_ku^k - f^Ac^2)\ =\ 0,
\\ \label{K9a}
J^A_ku^{Ak}&=& c^2\rh^A[1-(f^A)^2]\ =:\ c^2\rh^Aw^A\
=\ w^AN^A_ku^{Ak},
\\ \label{K9b}
1-w^A\ \geq\ 0.\hspace{-7.3cm}
\eey
By introducing the projectors
\bee{K15}
h^{Am}_l\ :=\ \delta^m_l - \frac{1}{c^2}u^{Am}u^{A}_l,\qquad
h^{m}_l\ :=\ \delta^m_l - \frac{1}{c^2}u^{m}u_l,
\ee
we obtain the following properties of the diffusion flux density:
\byy{L1}
J^{Am}h^k_m &=& J^{Ak}\ =\ N^{Am} h^k_m
\\ \label{L2}
J^{Am}h^{Ak}_m &=& \rh^Af^A(f^Au^{Ak}-u^k)
\\ \label{L3}
J^{Ak} &=& J^{Am}h^{Ak}_m + \rh^Aw^Au^{Ak}\ =\
J^{Am}h^{Ak}_m+w^AN^{Ak}                                                 
\\ \label{K8d1}
J^{Ak}{_{;k}} &=&(J^{Am}h^{Ak}_m)_{;k}
+(\rh^Aw^A){_{;k}}u^{Ak}+\rh^Aw^Au^{Ak}{_{;k}}.
\eey
According to \R{L1}$_2$, the diffusion flux density is that part of the mass flux density
which is perpendicular to the 4-velocity of the mixture. The diffusion flux density
vanishes in 1-component systems ($u^A_k\equiv u_k$) according to $f^A=f=1$ and \R{K8c}$_1$.

\section{The Energy-Momentum Tensor}
\subsection{Free and interacting components\label{FIC}}

The energy-momentum tensor $T^{Akl}$ of the $^A$-component consists of two parts
\bee{O1}
T^{Akl}\ =\ \st{0}{T}\!{^{Akl}} + \sum_B W^{Akl}_B,\quad W^{Bkl}_B\ =\ 0.
\ee
Here, $\st{0}{T}\!{^{Akl}}$ is the energy-momentum tensor of the free
$^A$-component, that is the case, if there are no interactions between the
$^A$-component and the other ones. $W^{Akl}_B$ describes the interaction
between the $^B$- and the $^A$-component.
The interaction between the external environment and the $^A$-component is given by the force density $k^{Al}$ which appears in the energy-momentum balance equation
\bee{O2}
T^{Akl}{_{;k}}\ =\ k^{Al}\ =\ \Omega^{Al}+\frac{1}{c^2}u^{Al}u^A_mk^{Am},
\qquad \Omega^{Al}u^A_l\ =\ 0,
\ee
and in the balance equations of
\byy{K13}
\mbox{energy:}\hspace{1.4cm}
u^A_lT^{Akl}{_{;k}} &=& u^A_lk^{Al}\ =:\ \Omega^A,
\\ \label{K14}
\mbox{and momentum:}\hspace{.5cm}
h^{Am}_lT^{Akl}{_{;k}} &=& h^{Am}_lk^{Al}\ =:\ \Omega^{Am}.
\eey
Consequently, the interaction of the $^A$-component with the other components
of the mixture modifies the energy-momentum tensor of the free $^A$-component.
Additionally, its interaction with the environment shows up in the source of
the energy-momentum balance. According to its definition, $T^{Akl}$ is the energy-momentum tensor of the "$^A$-component in the mixture".

\subsection{(3+1)-split}

The (3+1)-split of the energy-momentum tensor of the $^A$-component is
\bee{J1}
T^{Akl}\ =\ \frac{1}{c^2}e^Au^{Ak}u^{Al} + 
u^{Ak}p^{Al} +\frac{1}{c^2}q^{Ak}u^{Al} +t^{Akl}.
\ee
The (3+1)-components of the energy-momentum tensor are\footnote{the (3+1)-split
is made by taking the physical meaning of \R{J2} and \R{J3} into account, see \R{J6} to \R{J8a}}
\byy{J2}
e^A\ :=\ \frac{1}{c^2}T^{Ajm}u^A_ju^A_m ,\qquad
p^{Al}\ :=\ \frac{1}{c^2}h^{Al}_mT^{Ajm}u^A_j,
\\ \label{J3}
q^{Ak}\ :=\ h^{Ak}_jT^{Ajm}u^A_m,\qquad
t^{Akl}\ :=\ h^{Ak}_jT^{Ajm}h^{Al}_m,
\\ \label{J3a}
q^{Ak}u^A_k = 0,\ p^{Al}u^A_l = 0,\quad t^{Akl}u^A_k =0,\ t^{Akl}u^A_l = 0.
\eey

The (3+1)-split of tensors is a usual tool in relativistic continuum
physics. The (3+1)-components --generated by the split-- have physical
significance which originally is hidden in the unsplitted tensors. Thus, we
generate by (3+1)-splitting the following covariant quantities of the $^A$-component:
the energy density $e^A$,
the momentum flux density $p^{Al}$, the energy flux density $q^{Ak}$, the
stress tensor $t^{Akl}$

The symmetric part of the energy-momentum tensor \R{J1} is
\bee{J3b}
T^{A(kl)}\ =\ \frac{1}{c^2}e^Au^{Ak}u^{Al}
+\frac{1}{2c}u^{Ak}\Big(cp^{Al}+\frac{1}{c}q^{Al}\Big)
+\frac{1}{2c}\Big(cp^{Ak}+\frac{1}{c}q^{Ak}\Big)u^{Al}+t^{A(kl)},
\ee
and its anti-symmetric part is
\bee{J3c}
T^{A[kl]}\ =\ \frac{1}{2c}u^{Ak}\Big(cp^{Al}-\frac{1}{c}q^{Al}\Big)
-\frac{1}{2c}\Big(cp^{Ak}-\frac{1}{c}q^{Ak}\Big)u^{Al}+t^{A[kl]}.
\ee
The stress tensor is composed of the pressure $p^A>0,\wedge A$, and the viscous tensor
$\pi^{Akl}$
\bee{cT2}
t^{Akl}\ =\ -p^Ah^{Akl} + \pi^{Akl},\qquad t^{Ak}_k\ =\ -3p^A.
\ee

We now consider the physical dimensions of the introduced quantities\footnote{the bracket [$\boxtimes$] signifies the physical dimension of $\boxtimes$}.
According to \R{K15} and \R{K6b}$_1$, we have
\bee{J5}
[h^{Al}_m]\ =\ 1,\qquad [f^A]\ =\ 1.
\ee
By taking \R{cT2}, \R{J5}$_1$ and \R{J1} into account we obtain
\byy{J6}
 [t^{Akl}]\ =\ [p^{A}]\ =\ [\pi^{Akl}]\ =\ [e^A]\ =\ [q^{Ak}]\frac{s}{m}&=& [p^{Al}]\frac{m}{s},
\\ \nonumber
\mbox{pressure}\ =\ [p^{A}]\ =\ \frac{N}{m^2}\ =\ \frac{Nm}{m^3}
&=& \mbox{energy density}\ =\hspace{1.2cm} 
\\ \label{J7}
=\ \frac{kg\ m}{s^2}\frac{1}{m^2}\
=\ kg\frac{m}{s}\frac{1}{m^3}\frac{m}{s}&=&\mbox{momentum flux density},
\\ \label{J8}
[q^{Ak}]\ =\ [e^A]\frac{m}{s}\ =\ \frac{Nm}{m^3}\frac{m}{s}&=&
\mbox{energy flux density},
\\ \label{J8a}
[p^{Al}]\ =\ kg\frac{m}{s}\frac{1}{m^3}&=&\mbox{momentum density}.
\eey

The (3+1)-split \R{J1} of the energy-momentum tensor can be written in a more
compact form 
\byy{K11a}
T^{Akl}\ =\ \frac{1}{c^2}Q^{Ak}u^{Al} + \tau^{Akl},\hspace{4.5cm}
\\ \label{K11b}
u_lT^{Akl}\ =:\ Q^{Ak}\ =\ e^Au^{Ak} + q^{Ak},\qquad
h_l^mT^{Akl}\ =:\ \tau^{Akm}\ =\ u^{Ak}p^{Am} + t^{Akm}.
\eey
The energy-momentum tensor \R{K11a} is that of the $^A$-component in the
mixture, that means as dicussed in sect.\ref{FIC}, the  $^A$-component is not a free system and the
(3+1)-split-components $e^A, q^{Ak}, p^{Al}$ and $ t^{Akl}$ include the internal
interaction of the $^A$-component with all the other ones.

\subsection{Additivity}

We now consider the equivalent-system composed of the $Z$  components: that is the mixture which consists of these $Z$ interacting components. Because this interaction
is already taken into account by the (3+1)-split-components, the
energy-momentum tensors of the components are additive without
additional interaction terms. Consequently, the energy-momentum tensor ${\sf T}^{kl}$ of
the mixture is
\bey\nonumber
\mbox{\sf Setting II:}\hspace{9.5cm}
\\ \label{K11c}
{\sf T}^{kl}\ :=\ \frac{1}{c^2}Q^{k}u^{l} + \tau^{kl}  \st{\td}{=}\
\sum_A T^{Akl}\ =\
\sum_A\Big(\frac{1}{c^2}Q^{Ak}u^{Al} + \tau^{Akl}\Big).
\eey
Multiplication with $u_l$ results by use of \R{K6b}$_1$ and \R{K11b}$_2$ in
\bee{K11d}
Q^k\ =\ \sum_A\Big(Q^{Ak}f^A + \tau^{Akl}u_l\Big),
\ee
and by multiplication with $h_l^m$,  \R{K11c} results in
\byy{K15a}
\tau^{km} &=& \sum_A\Big(Q^{Ak}g^{Am} +
\tau^{Akl}h_l^m\Big),
\\ \label{K15a1}
g^{Am} &:=& \frac{1}{c^2}u^{Al}h_l^m\ =\ 
\frac{1}{c^2}(u^{Am}-f^Au^m)\ =\ \frac{1}{c^2\rh^A}J^{Am}.
\eey
For an 1-component system ($u^A_k\equiv u_k$), we obtain according to \R{K15a1}
$g^{Am}=g^m=0$ taking $f^A=f=0$ into account.

\subsection{(3+1)-components of the mixture\label{COMI}}

Starting with \R{K11b}, we obtain
\byy{K15a2}
Q^{Ak}u^A_k\ =\ c^2e^A,&\quad& Q^{Ak}h_k^{Am}\ =\ q^{Am},
\\ \label{K15a3}
\tau^{Akm}u^A_k\ =\ c^2p^{Am},&\quad&
\tau^{Akm}h_k^{Aj}\ =\ t^{Ajm}.
\eey
According to \R{K11a} and \R{K11c}$_1$, these relations are analogous for the mixture.
Consequently, from \R{K11d} follows
\bee{K15b}
Q^ku_k\ =:\ c^2{\sf e}\ =\ \sum_A\Big(Q^{Ak}f^Au_k +
\tau^{Akl}u_lu_k\Big),
\ee
resulting with \R{K15a2}$_1$ in the energy density of the mixture
\bee{K15c}
c^2{\sf e}\ =\ \sum_A\Big(c^2e^A(f^A)^2 + q^{Ak}f^Au_k +c^2p^{Al}f^Au_l +
t^{Akl}u_lu_k\Big).
\ee
From \R{K11d} follows
\bee{K15d}
Q^kh_k^m\ =:\ {\sf q}^m\ =\ \sum_A\Big(f^AQ^{Ak} h^m_k +
h^m_k\tau^{Akl}u_l\Big),
\ee
resulting with \R{K15a2}$_2$ in the energy flux density of the mixture
\bee{K15e}
{\sf q}^m\ =\ \sum_A\Big(c^2e^Af^Ag^{Am} + q^{Ak}f^Ah^m_k +
c^2p^{Al}g^{Am}u_l + t^{Akl}h^m_ku_l\Big).
\ee
From \R{K15a} follows
\bee{K15f}
\tau^{km}u_k\ =:\ c^2{\sf p}^m\ =\ \sum_A\Big(Q^{Ak}u_kg^{Am} +
\tau^{Akl}h^m_lu_k\Big),
\ee
resulting with \R{K15a3}$_1$ in the momentum density of the mixture
\bee{K15g}
c^2{\sf p}^m\ =\ \sum_A\Big(c^2e^Af^Ag^{Am} + q^{Ak}u_kg^{Am} +
c^2p^{Al}f^Ah^m_l + t^{Akl}h^m_lu_k\Big).
\ee
And from \R{K15a3}$_2$ follows finally
\bee{K15h}
\tau^{km}h_k^j\ =:\ {\sf t}^{jm}\ =\ \sum_A\Big(Q^{Ak}g^{Am}h^j_k +
\tau^{Akl}h^m_lh^j_k\Big)
\ee
which by taking \R{K11b} into account results in the stress tensor and the pressure of the
mixture
\byy{K15j}
{\sf t}^{jm} &=& \sum_A\Big(c^2e^Ag^{Aj}g^{Am} + q^{Ak}h^j_kg^{Am} +
c^2p^{Al}g^{Aj}h^m_l + t^{Akl}h^j_kh^m_l\Big),
\\ \nonumber
{\sf p} &=& -\frac{1}{3}{\sf t}^{jm}h_{jm}\ =\ 
-\frac{1}{3}\sum_A\Big(Q^{Ak}g^{Am}h^j_k + \tau^{Akl}h^m_lh^j_k\Big)h_{jm}\
=
\\  \nonumber
&\mbox{}&\hspace{2.1cm} =\ 
-\frac{1}{3}\sum_A\Big(\frac{1}{c^2}Q^{Ak}u^{Ap}h_{kp} + \tau^{Akl}h_{kl}\Big)\
=
\\ \label {K15j1}
&\mbox{}&\hspace{2.1cm}=\ 
-\frac{1}{3}\sum_A\Big(\frac{1}{c^2}( e^Au^{Ak} + q^{Ak})u^{Ap}h_{kp} +
(u^{Ak}p^{Al} + t^{Akl})h_{kl}\Big).
\eey

The additivity of the energy-momentum tensors \R{K11c}
results in \R{K15c}, \R{K15e}, \R{K15g} and \R{K15j}, relations which
express the (3+1)-components of the energy-momentum tensor of the
mixture as those of the components and their velocities
\byy{K15k}
\Big\{{\sf e},{\sf q}^k,{\sf p}^k,{\sf t}^{kl}\Big\} &=& F\Big(e^A,q^{Ak},p^{Ak},t^{Akl},
u^{Ak},\rh(\rh^A,u^A_k,u^k), u^k(\rh^A,u^A_k,\rh)\Big),
\\ \label{K15k1}
{\sf T}^{kl} &=& \frac{1}{c^2}{\sf e}u^{k}u^{l} + 
u^{k}{\sf p}^{l} +\frac{1}{c^2}{\sf q}^{k}u^{l} +{\sf t}^{kl},\qquad
{\sf t}^{kl}\ =\ -{\sf p}h^{kl}+^\dm\!\pi^{kl}.
\eey
The 4-velocity $u^k$ is given by \R{K5}$_4$.

The influence of the additivity of the energy-momentum tensors on the
balance equations of energy and momentum is investigated in the next section.

\section{Entanglement of Energy and Momentum Balances}

If the energy-momentum tensors of the $^A$-component and of the mixture
are $T^{Akl}$ and ${\sf T}^{kl}$, the energy and momentum balances are
according to the mixture axiom by use of \R{K13} and \R{K14}
\byy{K13z}
\mbox{energy:}\hspace{1.4cm}
u^A_lT^{Akl}{_{;k}}\ =\ \Omega^A,
&\quad& u_l{\sf T}^{kl}{_{;k}}\ =\ \Omega,
\\ \label{K14z}
\mbox{momentum:}\hspace{.5cm}
h^{Am}_lT^{Akl}{_{;k}}\ =\ \Omega^{Am},
&\quad& h^m_l{\sf T}^{kl}{_{;k}}\ =\ \Omega^m.
\eey
The balances \R{K13z}$_3$ and \R{K14z}$_3$ follow from \R{K13} and \R{K14}
by the mixture axiom.
Here, $\Omega^A$ and  $\Omega$ are the energy supplies, and
$\Omega^{Am}$ and  $\Omega^m$ the momentum supplies of
the $^A$-component and of the mixture.

The (3+1)-split of the divergence of the energy-momentum tensor of the
$^A$-component results by use of \R{K15}$_1$ in
\bee{K14a}
\delta^m_lT^{Akl}{_{;k}}\ =\ T^{Akm}{_{;k}}\ =\ h^{Am}_lT^{Akl}{_{;k}}+
\frac{1}{c^2}u^{Am}u^A_lT^{Akl}{_{;k}}.
\ee
If the component index $^A$ is cancelled in \R{K14a}, we obtain
the decomposition of the divergence of the energy-momentum tensor of the mixture.
Taking \R{K13z} and \R{K14z} into account, these divergences can be written as
\bee{K14d}
T^{Akm}{_{;k}}\ =\ \Omega^{Am} + \frac{1}{c^2}u^{Am}\Omega^A,\qquad
{\sf T}^{km}{_{;k}}\ =\ \Omega^{m} + \frac{1}{c^2}u^m\Omega.
\ee

The additivity of the energy-momentum tensors \R{K11c} results in the additivity of
the force densities\footnote{this is a strong argument for the validity of Setting II
\R{K11c}}
\bee{K14e} 
k^m\ =\ \Omega^{m} + \frac{1}{c^2}u^m\Omega\ =\
\sum_A\Big(\Omega^{Am} + \frac{1}{c^2}u^{Am}\Omega^A\Big)\ =\
\sum_A k^{Am}.
\ee
Taking \R{K13}$_2$ and \R{K14}$_2$ into account, we obtain by
multiplication of \R{K14e} with $u_m$, resp. with $h^p_m$,
\bee{K14f}
\Omega\ =\ \sum_A\Big(\Omega^{Am}u_m + f^A\Omega^A\Big),
\qquad
\Omega^p\ =\ \sum_A\Big(\Omega^{Am}h^p_m + 
g^{Ap}\Omega^A\Big).
\ee
Inserting \R{K13} and \R{K14}, we obtain in more detail
\byy{K14g}
u_l{\sf T}^{kl}{_{;k}}\ =\ \sum_A\Big\{\Big(h^{Am}_lu_m +
f^Au^A_l\Big)T^{Akl}{_{;k}}\Big\},
\\ \label{K14h}
h^p_l{\sf T}^{kl}{_{;k}}\ =\ \sum_A\Big\{\Big(h^{Am}_lh^p_m +
g^{Ap}u^A_l\Big)T^{Akl}{_{;k}}\Big\}.
\eey
As \R{K14f} indicates, the additivity of the energy-momentum
tensors causes
that the supplies of energy and momentum are entangled, expressed
by the inequalities
\bee{K16}
\sum_A f^A\Omega^A\ \neq\ \Omega,\qquad
\sum_A \Omega^{Am}h^p_m\ \neq\ \Omega^p.
\ee

Also if the total force density and the total momentum supply are zero,
\bee{K17}
{\sf T}^{kl}{_{;k}}\ =\ 0\quad\longrightarrow\quad\Omega_{\sf iso}\ =\ 0\
\wedge\ \Omega^m_{\sf iso}\ =\ 0,
\ee
we obtain according to \R{K14f}$_{1,2}$
\byy{K18}
\sum_A\Omega^{Am}_{\sf iso}u_m &=&
-\sum_Af^A\Omega^A_{\sf iso}\ \neq\ 0,
\\ \label{K19}
\sum_A\Omega^{Am}_{\sf iso}h^p_m &=& 
-\sum_Ag^{Ap}\Omega^A_{\sf iso}\ \neq\ 0.
\eey
As expected,  the supplies of energy and momentum remain entangled in a system
of vani\-shing total force and momentum densities. The entanglement vanishes for such isolated
systems for which the force and momentum supplies for all $^A$-components are zero.

\section{The Spin Tensor}
\subsection{(3+1)-split}

The (3+1)-split of the spin tensor of an $^A$-component is defined by inserting \R{K15}$_1$ into
\bee{+K19}
S^{Akab}\ =\ S^{Ampq}\delta_m^k\delta_p^a\delta_q^b.
\ee
Introducing the following covariant abreviations
\byy{K19b}
s^{Amj}\ :=\ S^{Akab}u^A_kh^{Am}_ah^{Aj}_b,\qquad 
s^{Amji}\ :=\ S^{Akab}h^{Am}_kh^{Aj}_ah^{Ai}_b
\\ \label{K19c}
\Xi^{Am}\ :=\ S^{Akab}u^A_ku^A_ah^{Am}_b,\qquad
\Xi^{Amj}\ :=\ S^{Akab}h^{Am}_ku^A_ah^{Aj}_b,
\eey
\R{+K19} results in
\bey\nonumber
S^{Akab}\ =\  -S^{Akba}\ =\ \hspace{10.5cm}
\\ \label{K19a}
=\ u^{Aa}\Big(\frac{1}{c^4}u^{Ak}\Xi^{Ab}+\frac{1}{c^2}\Xi^{Akb}\Big)
-u^{Ab}\Big(\frac{1}{c^4}u^{Ak}\Xi^{Aa}+\frac{1}{c^2}\Xi^{Aka}\Big)
+s^{Akab}+ \frac{1}{c^2}u^{Ak}s^{Aab}.\hspace{.2cm}
\eey
By \R{K19b} and \R{K19c} are introduced:
the spin density $s^{ab}$, the spin density
vector $\Xi^b$, the couple stress $s^{kab}$ and the spin stress $\Xi^{kb}$.

Analogously to \R{K11a} and \R{K11b}, a more compact form of the spin tensor is
\byy{K19d}
&&S^{Akab}\ =\ 2u^{A[a}L^{Akb]} +M^{Akab},
\\ \label{K19e}
 L^{Akb}\ :=\ \frac{1}{c^4}u^{Ak}\Xi^{Ab}+\frac{1}{c^2}\Xi^{Akb},\quad
&& M^{Akab}\ :=\ s^{Akab}+ \frac{1}{c^2}u^{Ak}s^{Aab}.
\eey
Taking \R{K19b} and \R{K19c} into account, we obtain
\bee{+K19f}
S^{Akab}u^A_a\ =\ c^2L^{Akb},\qquad S^{Akab}h_a^{Am}h_b^{An}\ =\ M^{Akmn},
\ee
expression which are needed for formulating the entropy identity below.

\subsection{Additivity}

Analogously to Setting II, we introduce the spin tensor of the mixture as the sum
of the spin tensors of the {$^A$-components}.
\bey\nonumber
\mbox{\sf Setting III:}
\\ \nonumber
{\sf S}^{kab} &:=& 2u^{[a}L^{kb]} +M^{kab}
\st{\td}{=}\
\sum_A S^{Akab}\ =\
\sum_A\Big(2u^{A[a}L^{Akb]} +M^{Akab}
\Big)\ =\hspace{.6cm} 
\\ \label{K19g}
&=& \sum_A\Big\{2u^{A[a}\Big(\frac{1}{c^4}u^{Ak}\Xi^{Ab]}+\frac{1}{c^2}\Xi^{Akb]}\Big)
+s^{Akab}+ \frac{1}{c^2}u^{Ak}s^{Aab}.
\Big\}.
\eey
According to the mixture axiom,  the spin tensor of the mixture is defined by \R{K19g}$_1$
as spin of an 1-component system resulting from \R{K19d} with $A\equiv 1\rightarrow blank$.

\subsection{(3+1)-components of the mixture}

From \R{K19g}$_1$ we obtain by taking the mixture axiom and \R{+K19f} into account
\byy{K19h}
{\sf S}^{kab}u_a &=& c^2L^{ab}\ =\
\sum_A\Big(2u^{A[a}L^{Akb]} +M^{Akab}\Big)u_a,
\\ \label{K19i}
{\sf S}^{kab}h_a^mh_b^n &=& M^{kmn}\ =\ 
 \sum_A\Big(2u^{A[a}L^{Akb]} +M^{Akab}\Big)h_a^mh_b^n.
\eey
The (3+1)-components of the spin tensor of the mixture result from \R{K19b} and \R{K19c}
using the mixture axiom
\byy{K19j}
s^{mj}\ =\ {\sf S}^{kab}u_kh^{m}_ah^{j}_b,\qquad 
s^{mji}\ =\ {\sf S}^{kab}h^{m}_kh^{j}_ah^{i}_b
\\ \label{K19k}
\Xi^{m}\ =\ {\sf S}^{kab}u_ku_ah^{m}_b,\qquad
\Xi^{mj}\ =\ {\sf S}^{kab}h^{m}_ku_ah^{j}_b,
\eey
and by inserting \R{K19g}$_4$ or \R{K19h} and \R{K19i} .

\subsection{Spin balance equation}

If there exists an external angular momentum density
\bee{K19l}
m^{ab}\ =\ -m^{ba},
\ee
a spin balance equation of each $^A$-component and of the mixture has to be taken into account
\bee{K19m}
S^{Akab}{_{;k}}\ 
=\ \frac{1}{c^2}m^{Aab},\qquad {\sf S}^{kab}{_{;k}}\ =\ \frac{1}{c^2}m^{ab}.
\ee
According to Setting III, 
\bee{K19n}
\sum_Am^{Aab}\ =\ m^{ab}
\ee
the additivity of the partial angular momenta is valid.

\section{Thermodynamics of Interacting Components\label{IC}}
\subsection{The entropy identity\label{IC1}}

Starting with the (3+1)-split of the entropy 4-vector and the entropy balance equation
\byy{+T2}
S^{Ak}\ =\ s^Au^{Ak} + s^{Ak}\ &\longrightarrow& S^{Ak}{_{;k}}\ =\ \sigma^A + \varphi^A,
\\ \label{+T2z}
s^A\ :=\ \frac{1}{c^2}S^{Ak}u^A_k,&\ & s^{Ak}\ :=\ S^{Am}h^{Ak}_m
\eey
we have to define the following four quantities in accordance with the balance equations of mass
\R{K2}$_2$,  of energy \R{K13}, of momentum \R{K14} and of spin \R{K19m}$_1$: 
the entropy density $s^A$, the entropy flux density $s^{Ak}$, the entropy production $\sigma^A$
and the entropy supply $\varphi^A$. Because there is no unequivocal entropy
\C{KEL} and consequently, also no unique entropy density, -flux, -production and-supply, we need a
tool which helps to restrict the arbitrariness for defining entropies. Such a tool is the {\em entropy identity} \C{RELTAG,MUBO} which is generated by adding suitable zeros to the entropy
\R{+T2}$_1$ which are related to the balances which are taking into account. These zeros are
generated by choosing the following expressions:
$N^{Ak},\ u^A_lT^{Akl},\ h^{Am}_lT^{Akl},\ u^A_aS^{Akab}, h^{Am}_ah^{An}_bS^{Akab}$. 
Consequently, the entropy identity is chosen according to \R{K2}, \R{K11b} and \R{+K19f}
\bey\nonumber
S^{Ak}&\equiv& s^Au^{Ak} + s^{Ak}
+\kappa^A\Big(N^{Ak}-\rh^Au^{Ak}\Big)+
\\ \nonumber
&&+\lambda^A\Big(u^A_lT^{Akl} - e^Au^{Ak}-q^{Ak}\Big)+
\\ \nonumber
&&+\lambda^A_m\Big(h^{Am}_lT^{Akl}-u^{Ak}p^{Am}-t^{Akm}\Big)+
\\  \nonumber
 &&+\Lambda ^A_{m}\Big(u^A_ah^{Am}_bS^{Akab}-\frac{1}{c^2}u^{Ak}\Xi^{Am}-\Xi^{Akm}\Big)+
\\ \label{+aT2}
&&+\Lambda ^A_{mn}\Big(h_a^{Am}h_b^{An}S^{Akab}-
 s^{Akmn}-\frac{1}{c^2}u^{Ak}s^{Amn}\Big).
\eey

The fields of Lagrange multipliers $\kappa^A,\ \lambda^A,\ \lambda^A_m$,
$\Lambda^A_{m}$ and $\Lambda^A_{mn}$ are quantities whose physical
meaning becomes clear in the course of the exploitation of the entropy identity.
Here, $\kappa^A$ and $\lambda^A$ are scalars, undefined for the present, and for
the likewise arbitrary quantities $\lambda^A_m$, $\Lambda^A_{m}$ and
$\Lambda^A_{ab}$, tensors of first and second order.
An identification of these Lagrange multipliers
is given below after the definitions of entropy flux density, entropy production density
and supply in section \ref{ACVA}.

The entropy identity \R{+aT2} depends on the balances which are taken into
consideration as constraints: the balances of mass, energy, momentum and spin.
The electro-magnetic field and quantum fields are included, if the energy-momentum tensor
and the spin tensor of these fields are inserted into \R{+aT2}.

Considering the third, the fourth and the fifth row of \R{+aT2}, we obtain that the velocity parts of
$\lambda^A_m$, $\Lambda^A_m$ and $\Lambda^A_{mn}$ can be set to zero according to
\R{K19b} and \R{K19c}.
The symmetric part of $\Lambda^A_{mn}$ does not contribute to the fifth row of \R{+aT2} and
therefore it is set to zero, too
\bee{*aT2}
\lambda^A_mh^{Am}_l\ =\ \lambda^A_l,\qquad
\Lambda^A_mh^{Am}_b\ =\ \Lambda^A_b,\qquad
\Lambda ^A_{mn}h_a^{Am}h_b^{An}\ =\ \Lambda ^A_{ab}\ =\ -\Lambda ^A_{ba}.
\ee

The entropy identity \R{+aT2} becomes by rearranging
\bey\nonumber
S^{Ak}&\equiv&
u^{Ak}\Big(s^A-\kappa^A\rh^A-\lambda^Ae^A-\lambda^A_mp^{Am}
-\Lambda ^A_{m}\frac{1}{c^2}\Xi^{Am}
-\Lambda ^A_{mn}\frac{1}{c^2}s^{Amn}\Big)+
\\ \nonumber
&&+s^{Ak}+\kappa^AN^{Ak}+\Big(\lambda^Au^A_l+\lambda^A_l\Big)T^{Akl}
-\lambda^Aq^{Ak}
-\lambda^A_lt^{Akl}+
\\ \label{+bT2}
&&+\Big(\Lambda ^A_{b}u^A_a+\Lambda^A_{ab}\Big)S^{Akab}
-\Lambda^A_m\Xi^{Akm}-\Lambda^A_{mn}s^{Akmn}.
\eey

This identity transforms into an other one by differentiation and by taking the 
balance equations of mass \R{K2}$_2$, of energy-momentum \R{O2}, of spin \R{K19m}$_1$ and
of entropy \R{+T2}$_2$ into account.
\bey\nonumber
S^{Ak}{_{;k}}&\equiv&
\Big[u^{Ak}\Big(s^A-\kappa^A\rh^A-\lambda^Ae^A
-\lambda^A_mp^{Am}-\Lambda ^A_{m}\frac{1}{c^2}\Xi^{Am}
-\Lambda ^A_{mn}\frac{1}{c^2}s^{Amn}
\Big)\Big]_{;k}+
\\ \nonumber
&&+s^{Ak}{_{;k}}+\kappa^A{_{;k}}N^{Ak}+\kappa^A\Big({^{(ex)}}\Gamma^A+^{(in)}\!\Gamma^A\Big)+
\\ \nonumber
&&+\Big(\lambda^Au^A_l+\lambda^A_l\Big)_{;k}T^{Akl}
+\Big(\lambda^Au^A_l+\lambda^A_l\Big)k^{Al}-
\\ \nonumber
&&-\Big(\lambda^Aq^{Ak}\Big){_{;k}}-\Big(\lambda^A_lt^{Akl}\Big){_{;k}}+
\\ \nonumber
&&+\Big(\Lambda ^A_{b}u^A_a+\Lambda^A_{ab}\Big)_{;k}S^{Akab}
+\Big(\Lambda ^A_{b}u^A_a+\Lambda^A_{ab}\Big)\frac{1}{c^2}m^{Aab}-
\\ \label{+cT2}
&&-\Big(\Lambda^A_m\Xi^{Akm}\Big)_{;k}-\Big(\Lambda^A_{mn}s^{Akmn}\Big)_{;k}\ 
=\ \sigma^A + \varphi^A .
\eey
Here, $\sigma^A$ is the entropy production  and $\varphi^A$ the
entropy supply of the $^A$-component. The identity
\R{+cT2} changes into the entropy production, if $s^A$, $s^{Ak}$
and $\varphi^A$ are specified below.

Rearranging the entropy identity results in
\bey\nonumber
S^{Ak}{_{;k}}&\equiv&
u^{Ak}{_{;k}}\Big(s^A-\kappa^A\rh^A-\lambda^Ae^A
-\lambda^A_mp^{Am}-\Lambda ^A_{m}\frac{1}{c^2}\Xi^{Am}
-\Lambda ^A_{mn}\frac{1}{c^2}s^{Amn}\Big)+
\\ \nonumber
&&+u^{Ak}\Big(s^A-\kappa^A\rh^A-\lambda^Ae^A
-\lambda^A_mp^{Am} -\Lambda ^A_{m}\frac{1}{c^2}\Xi^{Am}
-\Lambda ^A_{mn}\frac{1}{c^2}s^{Amn}\Big)_{;k}+
\\ \nonumber
&&+\Big(s^{Ak}-\lambda^Aq^{Ak}-\lambda^A_lt^{Akl}
-\Lambda^A_m\Xi^{Akm}-\Lambda^A_{mn}s^{Akmn}\Big)_{;k}+
\\ \nonumber
&&+\kappa^A{^{(ex)}}\Gamma^A+\Big(\lambda^Au^A_l+\lambda^A_l\Big)k^{Al}
+\Big(\Lambda ^A_{b}u^A_a+\Lambda^A_{ab}\Big)\frac{1}{c^2}m^{Aab}+
\\ \nonumber
&&+\kappa^A{_{;k}}N^{Ak}+\kappa^A{^{(in)}}\!\Gamma^A+
\Big(\lambda^Au^A_l+\lambda^A_l\Big){_{;k}}T^{Akl}
+\Big(\Lambda ^A_{b}u^A_a+\Lambda^A_{ab}\Big)_{;k}S^{Akab}\ =\ \hspace{.5cm}
\\ \label{+T4}
&&=\ \sigma^A + \varphi^A .
\eey

Now we look for terms of the fifth row of \R{+T4} which fit into the first three rows
of \R{+T4}. The shape of these terms is $[u^{Ak}{_{;k}}{\sf scalar}/ u^{Ak}{\sf scalar}{_{;k}}]$ according to the first two rows of \R{+T4} and $[\Psi^{Ak}{_{;k}}\
(\Psi^{Ak}u^A_k=0)]$ according to the third row. None of the seven terms of the
fourth and fifth row of \R{+T4} have this
shape, but inserting the energy-momentum tensor and the spin tensor
into the fifth row of \R{+T4} may generate such
terms.

The third term of the fifth row of \R{+T4} becomes
\bey\nonumber
(\lambda^A u^A_l)_{;k}T^{Akl}\hspace{-.3cm} &=&\hspace{-.3cm}
\Big(\lambda^A{_{;k}}u^A_l +
\lambda^Au^A{_{l;k}}\Big)
\Big(\frac{1}{c^2}e^Au^{Ak}u^{Al}+
u^{Ak}p^{Al}+\frac{1}{c^2}q^{Ak}u^{Al}+t^{Akl}\Big)=
\\ \label{+T9a3}
&=&\hspace{-.3cm} \lambda{^A}{_{;k}}u^{Ak}e^A
+\lambda^A u^A{_{l;k}} u^{Ak}p^{Al}
+\lambda^A{_{;k}}q^{Ak}
-\U{p^A\lambda^Au^{Ak}{_{;k}}}
+\lambda^Au^{A}{_{l;k}}\pi^{Akl},\hspace{.9cm}
\\ \label{+Y14}
\lambda^A{_{l;k}}T^{Akl}\hspace{-.3cm}&=&\hspace{-.3cm}\lambda^A{_{l;k}} \Big(\frac{1}{c^2}e^Au^{Ak}u^{Al}+
u^{Ak}p^{Al}+\frac{1}{c^2}q^{Ak}u^{Al}+t^{Akl}\Big)
\eey
Summing up \R{+T9a3} and \R{+Y14} results in\footnote{the signs
$\U{\boxdot}$,
$\overbrace{\boxdot}$,
$\underbrace{\boxdot}$
and $\widetilde{\boxdot}$
mark terms which are related to each other in the sequel}
\bey\nonumber
\Big(\lambda^A u^A_l+\lambda^A_l\Big)_{;k}T^{Akl}\ 
=\ \lambda{^A}{_{;k}}\Big(q^{Ak}+e^Au^{Ak}\Big)
+\ \lambda^Au^A{_{l;k}}\Big(\pi^{Akl}+u^{Ak}p^{Al}\Big)-
\\ \label{+Y15}
-\U{p^A\lambda^Au^{Ak}{_{;k}}}
+\ \lambda^A{_{l;k}} \Big(\frac{1}{c^2}e^Au^{Ak}u^{Al}
+u^{Ak}p^{Al}+\frac{1}{c^2}q^{Ak}u^{Al}+t^{Akl}\Big).\hspace{.3cm}
\eey
Evidently, the term $-\U{p^A\lambda^Au^{Ak}{_{;k}}}$ belongs to the first row of \R{+T4}.
After having inserted the underlined term of \R{+Y15}, 
the first two rows of \R{+T4} become\footnote{\ $^\td$ is the
"component time derivative" $\st{\td}{\boxplus}{^A}:=\boxplus^A{_{;k}}u^{Ak}$}
\bey\nonumber
&&u^{Ak}{_{;k}}\Big(s^A  -\kappa^A\rh^A-\lambda^A e^A - \U{p^A\lambda^A}
- \lambda^A_mp^{Am}-\Lambda ^A_{m}\frac{1}{c^2}\Xi^{Am}
-\Lambda ^A_{mn}\frac{1}{c^2}s^{Amn}\Big)+
\\ \nonumber
&&+\Big(s^A-\kappa^A\rh^A -\lambda^A e^A - \underbrace{p^A\lambda^A}
- \lambda^A_mp^{Am}-\Lambda ^A_{m}\frac{1}{c^2}\Xi^{Am}
-\Lambda ^A_{mn}\frac{1}{c^2}s^{Amn}\Big){^\td}
+\underbrace{(p^A\lambda^A){^\td}}=\hspace{.7cm}
\\ \nonumber
&&=\ 
\Big[u^{Ak}\Big(s^A-\kappa^A\rh^A -\lambda^A e^A-p^A\lambda^A
- \lambda^A_mp^{Am}
-\Lambda ^A_{m}\frac{1}{c^2}\Xi^{Am}
-\Lambda ^A_{mn}\frac{1}{c^2}s^{Amn}\Big)\Big]_{;k}+
\\ \label{+T10}
&&+\widetilde{(p^A\lambda^A){^\td}}.
\eey
Thus, a rearranging of the entropy identity \R{+T4} results  by taking
\R{+T10} into account
\bey\nonumber
&&S^{Ak}{_{;k}}\ \equiv
\\ \nonumber
&&\Big[u^{Ak}\Big(s^A-\kappa^A\rh^A -\lambda^A e^A
-p^A\lambda^A
- \lambda^A_mp^{Am}-\Lambda ^A_{m}\frac{1}{c^2}\Xi^{Am}
-\Lambda ^A_{mn}\frac{1}{c^2}s^{Amn}\Big)\Big]_{;k}+\hspace{.4cm}
\\ \nonumber
&&+\widetilde{(p^A\lambda^A){^\td}}+\Big(s^{Ak}-\lambda^Aq^{Ak}-\lambda^A_lt^{Akl}
-\Lambda^A_m\Xi^{Akm}-\Lambda^A_{mn}s^{Akmn}\Big){_{;k}}+
\\ \nonumber
&&+\kappa^A{^{(ex)}}\Gamma^A+\Big(\lambda^Au^A_l+\lambda^A_l\Big)k^{Al}
+\Big(\Lambda ^A_{b}u^A_a+\Lambda^A_{ab}\Big)\frac{1}{c^2}m^{Aab}+
\\ \nonumber
&&+\kappa^A{_{;k}}N^{Ak}+\kappa^A{^{(in)}}\!\Gamma^A
+\Big(\Lambda ^A_{b}u^A_a+\Lambda^A_{ab}\Big)_{;k}S^{Akab}+
\\ \nonumber
&&+\lambda{^A}{_{;k}}\Big(q^{Ak}+e^Au^{Ak}\Big)
+\ \lambda^Au^A{_{l;k}}\Big(\pi^{Akl}+u^{Ak}p^{Al}\Big)+
\\ \label{T10}
&&+\ \st{\td}{\lambda}{^A_l} \Big(\frac{1}{c^2}e^Au^{Al}+p^{Al}\Big)
+\lambda^A{_{l;k}}\Big(\frac{1}{c^2}q^{Ak}u^{Al}+t^{Akl}\Big)\ =\ \sigma^A + \varphi^A .\hspace{.3cm}
\eey

The third term of the fourth row of \R{T10} results in
\bey\nonumber
\Big(\Lambda ^A_{b}u^A_a+\Lambda^A_{ab}\Big)_{;k}S^{Akab}\ =\hspace{8cm}
\\ \label{T10+}
=\Big(\Lambda ^A_{b;k}u^A_a +\Lambda ^A_{b}u^A_{a;k}+\Lambda^A_{ab;k}\Big)
\Big(u^{Aa}L^{Akb}-u^{Ab}L^{Aka}+M^{Akab}\Big).
\eey
If \R{*aT2} and \R{K19e} are taken into account, these nine terms are:
\byy{+T10+}
\Lambda ^A_{b;k}c^2L^{Akb}&=&
\Lambda ^A_{b;k}\Big(\frac{1}{c^2}u^{Ak}\Xi^{Ab}+\Xi^{Akb}\Big),
\\ \label{T10+a}
\Lambda ^A_{b}\ub{u^A_{a;k}}\ub{u^{Aa}}L^{Akb}&=&0,
\\ \label{T10+b}
\Lambda^A_{ab;k}u^{Aa}L^{Akb}&=&\Lambda^A_{ab;k}u^{Aa}
\Big(\frac{1}{c^4}u^{Ak}\Xi^{Ab}+\frac{1}{c^2}\Xi^{Akb}\Big),
\\ \label{T10+c}
-\Lambda ^A_{b;k}u^A_au^{Ab}L^{Aka}&=&-\Lambda ^A_{b;k}\ub{u^A_a}u^{Ab}
\Big(\frac{1}{c^4}u^{Ak}\ub{\Xi^{Aa}}+\frac{1}{c^2}\ub{\Xi^{Aka}}\Big)\ =\ 0,
\\ \label{T10+d}
-\ub{\Lambda ^A_{b}}u^A_{a;k}\ub{u^{Ab}}L^{Aka}&=&0,
\\ \label{T10+e}
-\Lambda^A_{ab;k}u^{Ab}L^{Aka}&=&-\Lambda^A_{ab;k}u^{Ab}
\Big(\frac{1}{c^4}u^{Ak}\Xi^{Aa}+\frac{1}{c^2}\Xi^{Aka}\Big),
\\ \label{T10+f}
\Lambda ^A_{b;k}\ub{u^A_a}\ub{M^{Akab}}&=&0,
\\ \label{T10+g}
\Lambda ^A_{b}u^A_{a;k}M^{Akab}&=&\Lambda ^A_{b}u^A_{a;k}
\Big(s^{Akab}+ \frac{1}{c^2}u^{Ak}s^{Aab}\Big),
\\ \label{T10+h}
\Lambda^A_{ab;k}M^{Akab}&=&\Lambda^A_{ab;k}
\Big(s^{Akab}+ \frac{1}{c^2}u^{Ak}s^{Aab}\Big).
\eey
Rearranging of \R{+T10+} to \R{T10+h} results in:
\bey\nonumber
\mbox{\R{+T10+} and \R{T10+g}:}\hspace{10cm}
\\ \label{T10+i}
\st{\td}{\Lambda}{_b^A}\frac{1}{c^2}\Xi^{Ab}+\Lambda ^A_{b;k}\Xi^{Akb}
-\Lambda ^A_{b}u^A_a\Big(s^{Akab}{_{;k}}+ \frac{1}{c^2}\st{\td}{s}\!{^{Aab}}\Big),
\\ \nonumber
\mbox{\R{T10+b}, \R{T10+e} and \R{T10+h}:}\hspace{8.8cm}
\\ \label{T10+j}
\st{\td}{\Lambda}\!{^A_{ab}}\Big(\frac{1}{c^4}u^{A[a}\Xi^{Ab]}+\frac{1}{c^2}s^{Aab}\Big)
+\Lambda}{^A_{ab;k}\Big(\frac{1}{c^2}u^{A[a}\Xi^{Akb]}+s^{Akab}\Big).
\eey

A comparison of \R{T10+i} and  \R{T10+j} with the first two rows of \R{T10}
demonstrates that a term which fits into these rows does not appear in \R{T10+i} and  \R{T10+j}.
Thus by taking \R{+T10} into account, a rearranging of the entropy identity \R{+T4} results in
\bey\nonumber
&S^{Ak}{_{;k}}& \hspace{-.3cm}\equiv
\\ \nonumber
&&\Big[u^{Ak}\Big(s^A-\kappa^A\rh^A -\lambda^A e^A
-p^A\lambda^A
- \lambda^A_mp^{Am}-\Lambda ^A_{m}\frac{1}{c^2}\Xi^{Am}
-\Lambda ^A_{mn}\frac{1}{c^2}s^{Amn}\Big)\Big]_{;k}+\hspace{.5cm}
\\ \nonumber
&&+\widetilde{(p^A\lambda^A){^\td}}+\Big(s^{Ak}-\lambda^Aq^{Ak}-\lambda^A_lt^{Akl}
-\Lambda^A_m\Xi^{Akm}-\Lambda^A_{mn}s^{Akmn}\Big)_{;k}+
\\ \nonumber
&&+\kappa^A{^{(ex)}}\Gamma^A+\Big(\lambda^Au^A_l+\lambda^A_l\Big)k^{Al}
+\Big(\Lambda ^A_{b}u^A_a+\Lambda^A_{ab}\Big)\frac{1}{c^2}m^{Aab}+
\\ \nonumber
&&+\kappa^A{_{;k}}N^{Ak}+\kappa^A{^{(in)}}\!\Gamma^A
+\lambda{^A}{_{;k}}\Big(q^{Ak}+e^Au^{Ak}\Big)
+\ \lambda^Au^A{_{l;k}}\Big(\pi^{Akl}+u^{Ak}p^{Al}\Big)+
\\ \nonumber
&&+\ \st{\td}{\lambda}{^A_l} \Big(\frac{1}{c^2}e^Au^{Al}+p^{Al}\Big)
+\lambda^A{_{l;k}}\Big(\frac{1}{c^2}q^{Ak}u^{Al}+t^{Akl}\Big)+
\\ \nonumber
&&+\st{\td}{\Lambda}{_b^A}\frac{1}{c^2}\Xi^{Ab}+\Lambda ^A_{b;k}\Xi^{Akb}
-\Lambda ^A_{b}u^A_a\Big(s^{Akab}{_{;k}}+ \frac{1}{c^2}\st{\td}{s}\!{^{Aab}}\Big)+
\\ \nonumber
&&+\st{\td}{\Lambda}\!{^A_{ab}}\Big(\frac{1}{c^4}u^{A[a}\Xi^{Ab]}+\frac{1}{c^2}s^{Aab}\Big)
+\Lambda}{^A_{ab;k}\Big(\frac{1}{c^2}u^{A[a}\Xi^{Akb]}+s^{Akab}\Big)\ =
\\ \label{T10y}
&&=\ \sigma^A + \varphi^A .
\eey
This entropy identity is incomplete: the multi-temperature relaxation is missing which is gene\-rated by the
different partial temperatures of the components of the mixture. Because of lucidity, the
treatment of multi-temperature relaxation is postponed and will be considered below in sect.\ref{PT}.
In the next section, we now specify $s^A$, $s^{Ak}$,  $\varphi^A$ and $\sigma^A$.

\subsection{Exploitation of the entropy identity\label{EXEI}}
\subsubsection{Entropy density, Gibbs and Gibbs-Duhem equations}

We now define the {\em entropy rest density} $s^A$ according to the first row of \R{T10y}
\bey\nonumber
\mbox{\sf Setting IV:}\hspace{11cm}
\\ \label{T10a}
s^A\ \st{\td}{=}\ \kappa^A\rh^A +\lambda^A e^A+p^A\lambda^A
+\lambda^A_lp^{Al}+\Lambda ^A_{m}\frac{1}{c^2}\Xi^{Am}
+\Lambda ^A_{mn}\frac{1}{c^2}s^{Amn},
\eey
resulting in the specific rest entropy
\bee{T10a1}
\frac{s^A}{\rh}\ =\ \kappa^A\frac{\rh^A}{\rh}
+\lambda^A\frac{e^A}{\rh}
+p^A\lambda^A\frac{1}{\rh}
+\lambda^A_l\frac{p^{Al}}{\rh}
+\Lambda ^A_{m}\frac{1}{c^2}\frac{\Xi^{Am}}{\rh}
+\Lambda ^A_{mn}\frac{1}{c^2}\frac{s^{Amn}}{\rh}.
\ee
A non-equilibrium state space --which is spanned by the independent variables--
contains beside $\rh^A$, $\rh$, $e^A$ the spin variables $\Xi^{Am}$ and $s^{Amn}$
and additionally $p^{Al}$ which extends
the state space in the sense of Extended Thermodynamics\footnote{If the energy-momentum tensor is presupposed to be symmetric --consequently
$p^{Al}=(1/c^2)q^{Al}$ is valid according to \R{J3c}-- the momentum density is
replaced by energy flux density which in  non-relativistic Extended Thermodynamics is set as a
non-equilibrium variable, even if the stress tensor is non-symmetric.}
\C{JOU,MUE}. Consequently, we choose the {\em state space} \C{SCH}
\bee{Y2}
{\sf z}^A\ =\ \Big(c^A,\frac{1}{\rh},\frac{e^A}{\rh},\frac{p^{Al}}{\rh},
\frac{\Xi^{Am}}{\rh},\frac{s^{Amn}}{\rh}\Big),
\qquad c^A\ :=\ \frac{\rh^A}{\rh}.
\ee

The corresponding {\em Gibbs equation} according to \R{T10a1} and \R{Y2} is
\bey\nonumber
\Big(\frac{s^A}{\rh}\Big)^\td\ =\
\kappa^A\st{\td}{c}{^A}
+\lambda^A \Big(\frac{e^A}{\rh}\Big)^\td
+p^A\lambda^A\Big(\frac{1}{\rh}\Big)^\td
+\lambda^A_l\Big(\frac{p^{Al}}{\rh}\Big)^\td+
\\ \label{Y3}
+\Lambda ^A_{m}\frac{1}{c^2}\Big(\frac{\Xi^{Am}}{\rh}\Big)^\td
+\Lambda ^A_{mn}\frac{1}{c^2}\Big(\frac{s^{Amn}}{\rh}\Big)^\td
\eey

Differentiation of \R{T10a1} results in the {\em Gibbs-Duhem equation} by taking
\R{Y3} into account
\bee{Y4}
0\ =\ \st{\td}{\kappa}\!{^A}c^A
+\st{\td}{\lambda}\!{^A}\frac{e^A}{\rh}
+(p^A\lambda^A)^\td\frac{1}{\rh}
+\st{\td}{\lambda}{^A_l}\frac{p^{Al}}{\rh}
+\st{\td}{\Lambda}\!{^A_{m}}\frac{1}{c^2}\frac{\Xi^{Am}}{\rh}
+\st{\td}{\Lambda}\!{^A_{mn}}\frac{1}{c^2}\frac{s^{Amn}}{\rh},
\ee
resulting in
\bee{Y4a}
\widetilde{(p^A\lambda^A)^\td}\ =\
- \st{\td}{\kappa}\!{^A} \rh^A
-\st{\td}{\lambda}\!{^A} e^A
-\st{\td}{\lambda}{^A_l}p^{Al}
-\st{\td}{\Lambda}\!{^A_{m}}\frac{1}{c^2}\Xi^{Am}
-\st{\td}{\Lambda}\!{^A_{mn}}\frac{1}{c^2}s^{Amn}.
\ee
Taking \R{Y4a} and \R{T10a} into account, the entropy identity \R{T10y} becomes
\bey\nonumber
&S^{Ak}{_{;k}}&\hspace{-.3cm}\equiv
-\ub{\st{\td}{\kappa}\!{^A}\rh^A}
-\U{\st{\td}{\lambda}\!{^A} e^A}
-\widehat{\st{\td}{\lambda}{^A_l}p^{Al}}
-\widetilde{\st{\td}{\Lambda}\!{^A_{m}}\frac{1}{c^2}\Xi^{Am}}
-\overbrace{\st{\td}{\Lambda}\!{^A_{mn}}\frac{1}{c^2}s^{Amn}} +
\\ \nonumber
&&+\ \Big(\ s^{Ak}
-\lambda^Aq^{Ak}
- \lambda^A_lt^{Akl}
-\Lambda^A_m\Xi^{Akm}-\Lambda^A_{mn}s^{Akmn}\Big)_{;k}+
\\ \nonumber
&&+\kappa^A{^{(ex)}}\Gamma^A+\Big(\lambda^Au^A_l+\lambda^A_l\Big)k^{Al}
+\Big(\Lambda ^A_{b}u^A_a+\Lambda^A_{ab}\Big)\frac{1}{c^2}m^{Aab}+
\\ \nonumber
&&+\ub{\kappa^A{_{;k}}N^{Ak}}+\kappa^A{^{(in)}}\!\Gamma^A
+\U{\lambda{^A}{_{;k}}}\Big(q^{Ak}+\U{e^Au^{Ak}}\Big)
+\ \lambda^Au^A{_{l;k}}\Big(\pi^{Akl}+u^{Ak}p^{Al}\Big)+
\\ \nonumber
&&+\ \widehat{\st{\td}{\lambda}{^A_l}} \Big(\frac{1}{c^2}e^Au^{Al}+
\widehat{p^{Al}}\Big)
+\lambda^A{_{l;k}}\Big(\frac{1}{c^2}q^{Ak}u^{Al}+t^{Akl}\Big)+
\\ \nonumber
&&+\widetilde{\st{\td}{\Lambda}{_b^A}\frac{1}{c^2}\Xi^{Ab}}+\Lambda ^A_{b;k}\Xi^{Akb}
-\Lambda ^A_{b}u^A_a\Big(s^{Akab}{_{;k}}+ \frac{1}{c^2}\st{\td}{s}\!{^{Aab}}\Big)+
\\ \nonumber
&&+\overbrace{\st{\td}{\Lambda}\!{^A_{ab}}}\Big(\frac{1}{c^4}u^{A[a}\Xi^{Ab]}
+\overbrace{\frac{1}{c^2}s^{Aab}}\Big)
+\Lambda}{^A_{ab;k}\Big(\frac{1}{c^2}u^{A[a}\Xi^{Akb]}+s^{Akab}\Big)\ =
\\ \label{T10yy}
&&=\ \sigma^A + \varphi^A .
\eey
The marked terms cancel each other.

Taking \R{K8c}$_2$ and \R{L3}$_2$ into account, we consider
\bey\nonumber
0 &=&\ub{\kappa^A{_{;k}}N^{Ak}} -\ub{\st{\td}{\kappa}\!{^A}\rh^A}\ =\
\kappa^A{_{;k}}\Big(N^{Ak}-N^{Ak}\Big)\ =\
\\ \nonumber
&=&\kappa^A{_{;k}}\Big(J^{Ak}+\rh^Af^Au^k-J^{Ak}-\rh^Af^Au^k\Big)\ =\
\kappa^A{_{;k}}\Big(J^{Ak}-J^{Ak}\Big)\ =\
\\ \nonumber
&=&\kappa^A{_{;k}}J^{Ak}
-\kappa^A{_{;k}}\Big(J^{Am}h^{Ak}_m + w^AN^{Ak}\Big)\ =\
\\ \label{Y15w}
&=&\kappa^A{_{;k}}\Big(\ub{J^{Ak}-w^AN^{Ak}}_{J^{Am}h^{Ak}_m}\Big)
-\Big(\kappa^AJ^{Am}h^{Ak}_m\Big)_{;k}+\kappa^A\Big(J^{Am}h^{Ak}_m\Big)_{;k}
\eey
This zero contains the diffusion flux which does not appear up to here in the entropy
identity \R{+aT2}. That means, the diffusion is missing in \R{T10yy}, and we will not ignore
the underbraced terms in \R{Y15w}$_1$, but we insert \R{Y15w}$_3$ into \R{T10yy}.
Consequently, the entropy identity results in
\bey\nonumber
S^{Ak}{_{;k}}\ \equiv\hspace{-.5cm} 
&& \Big(\ s^{Ak}
-\lambda^Aq^{Ak}
- \lambda^A_lt^{Akl}
-\Lambda^A_m\Xi^{Akm}-\Lambda^A_{mn}s^{Akmn}
-\kappa^AJ^{Am}h^{Ak}_m\Big)_{;k}
\\ \nonumber
&&+\kappa^A{^{(ex)}}\Gamma^A+\Big(\lambda^Au^A_l+\lambda^A_l\Big)k^{Al}
+\Big(\Lambda ^A_{b}u^A_a+\Lambda^A_{ab}\Big)\frac{1}{c^2}m^{Aab}+
\\ \nonumber
&&+\kappa^A{_{;k}}J^{Am}h^{Ak}_m
+\kappa^A\Big[{^{(in)}}\!\Gamma^A+\Big(J^{Am}h^{Ak}_m\Big)_{;k}\Big]+
\\ \nonumber
&&+\lambda{^A}{_{;k}}q^{Ak}
+\ \lambda^Au^A{_{l;k}}\Big(\pi^{Akl}+u^{Ak}p^{Al}\Big)+
\\ \nonumber
&&+\ \lambda^A{_{l;k}} \Big(\frac{1}{c^2}e^Au^{Ak}u^{Al}
+\frac{1}{c^2}q^{Ak}u^{Al}+t^{Akl}\Big)+
\\ \nonumber
&&+\Lambda ^A_{b;k}\Xi^{Akb}
-\Lambda ^A_{b}u^A_a\Big(s^{Akab}{_{;k}}+ \frac{1}{c^2}\st{\td}{s}\!{^{Aab}}\Big)+
\\ \nonumber
&&+\st{\td}{\Lambda}\!{^A_{ab}}\frac{1}{c^4}u^{A[a}\Xi^{Ab]}
+\Lambda}{^A_{ab;k}\Big(\frac{1}{c^2}u^{A[a}\Xi^{Akb]}+s^{Akab}\Big)\ =
\\ \label{Y15v}
&=& \sigma^A + \varphi^A .
\eey
We now specify the entropy flux density $s^{Ak}$ and the entropy supply $\varphi^A$
in the next section.

\subsubsection{Entropy flux, --supply and --production}

According to the first row of \R{Y15v}, we define the {\em entropy flux density}
\bey\nonumber
\mbox{\sf Setting V:}\hspace{8cm}
\\ \label{T7} 
s^{Ak}\ \st{\td}{=}\ \lambda^Aq^{Ak}
+\lambda^A_lt^{Akl}
+\Lambda^A_m\Xi^{Akm}+\Lambda^A_{mn}s^{Akmn}
+\kappa^AJ^{Am}h^{Ak}_m.
\eey

We now split the entropy identity \R{Y15v} into the entropy production and the entropy
supply.  For this end, we need a criterion to distinguish between entropy production and supply.
Such a criterion is clear for discrete systems: a local isolation suppresses the entropy
supply but not the entropy production. Isolation means: the second row in \R{Y15v}
vanishes, if the $^A$-component is isolated from the exterior of the mixture. Consequently,
we define the {\em entropy supply} as follows
\bey\nonumber
\mbox{\sf Setting VI:}\hspace{8cm}
\\ \label{T8}
\varphi^A\ \st{\td}{=}\
\kappa^A{^{(ex)}}\Gamma^A+\Big(\lambda^Au^A_l+\lambda^A_l\Big)k^{Al}
+\Big(\Lambda ^A_{b}u^A_a+\Lambda^A_{ab}\Big)\frac{1}{c^2}m^{Aab},
\eey
with the result that the entropy identity \R{Y15v} transfers into the entropy production density by
taking \R{T7} and \R{T8} into account
\bey\nonumber
\sigma^A\ =\ \hspace{-.5cm}
&&+\kappa^A{_{;k}}J^{Am}h^{Ak}_m
+\kappa^A\Big[{^{(in)}}\!\Gamma^A+\Big(J^{Am}h^{Ak}_m\Big){_{;k}}\Big]+
\\ \nonumber
&&+\lambda{^A}{_{;k}}q^{Ak}
+\ \lambda^Au^A{_{l;k}}\Big(\pi^{Akl}+u^{Ak}p^{Al}\Big)+
\\ \nonumber
&&+\ \lambda^A{_{l;k}} \Big(\frac{1}{c^2}e^Au^{Ak}u^{Al}
+\frac{1}{c^2}q^{Ak}u^{Al}+t^{Akl}\Big)+
\\ \nonumber
&&+\Lambda ^A_{b;k}\Xi^{Akb}
-\Lambda ^A_{b}u^A_a\Big(s^{Akab}{_{;k}}+ \frac{1}{c^2}\st{\td}{s}\!{^{Aab}}\Big)+
\\ \label{Y15w}
&&+\st{\td}{\Lambda}\!{^A_{ab}}\frac{1}{c^4}u^{A[a}\Xi^{Ab]}
+\Lambda}{^A_{ab;k}\Big(\frac{1}{c^2}u^{A[a}\Xi^{Akb]}+s^{Akab}\Big).
\eey

As expected, the entropy production is composed of terms which are a product of
"forces" and "fluxes" as in the non-relativistic case\footnote{The mass production
$^{(in)}\Gamma^A$ due to chemical reactions can be expressed by the time rate of
the reaction velocity, see \R{Z13c} in sect.\ref{SE}.}. The expressions
$s^A,\ s^{Ak},\ \varphi^A$ and $\sigma^A$ contain Lagrange multipliers which are
introduced for formulating the entropy identity \R{+aT2} playing up to here the role of
place-holders. Their physical meaning is discussed in the next section.

\subsection{Fields of Lagrange multipliers\label{ACVA}}

From non-relativistic physics, we know the physical dimensions of the entropy density and the
entropy flux density by taking \R{J6} and \R{J8} into account
\bee{T2w}
[s^A]\ =\ [e^A]\frac{1}{K}\ =\ \frac{Nm}{m^3}\frac{1}{K},\quad
[s^{Ak}]\ =\ [q^{Ak}]\frac{1}{K}\ =\
 \frac{Nm}{m^3}\frac{m}{s}\frac{1}{K}.
\ee
According to \R{T10a}, we have the following equation of physical dimensions
\bee{Z1}
[s^{A}]\ =\ [\lambda^A][e^A].
\ee
Taking \R{T2w}$_1$ and \R{J6} into account, we obtain
\bee{Z2}
\frac{N}{m^2}\frac{1}{K}\ =\ [\lambda^A] \frac{N}{m^2}
\quad\longrightarrow\quad
[\lambda^A]\ =\ \frac{1}{K},
\ee
that means, $\lambda^A$ is a reciprocal temperature belonging to the
$^A$-component. Therefore, we accept the following
\bey\nonumber
\mbox{\sf Setting VII:}\hspace{5cm}
\\ \label{Z3}
\lambda^A\ \st{\td}{=}\ \frac{\nu^A}{\Theta^A},\hspace{3cm}
\eey
with the partial temperature  $\Theta^A$ of the $^A$-component\footnote{This
temperature is a non-equilibrium one, the contact temperature
\C{MUTEMP,MUTEMP1,MU18a}
which should not be confused with the thermostatic equilibrium temperature
$\Theta^A_{eq}=T,\ \wedge A$.} and a scalar $\nu^A$ which is suitably chosen below..

According to \R{T7} , we have the following equation of physical dimensions
\bee{Z4}
[s^{Ak}]\ =\ [\kappa^A][J^{Am}][h_m^{Ak}].
\ee
Taking \R{T2w}$_2$, \R{K8c}$_2$ and \R{J5}$_1$ into account, we obtain 
\bee{Z5}
\frac{N}{ms}\frac{1}{K}\ =\ [\kappa^A]\frac{kg}{m^3}\frac{m}{s}1
\quad\longrightarrow\quad
[\kappa^A]\ =\ \frac{m^2}{s^2}\frac{1}{K}.
\ee
We know from the non-relativistic Gibbs equation that the chemical potentials $\mu^A$
have the physical dimension of the specific energy $e^A/\rh^A$
\bee{Z5a}
[\mu^A]\ = \frac{[e^A]}{[\rh^A]}\ =\ \frac{N}{m^2}\frac{m^3}{kg}\ =\
\frac{kg\ m}{s^2}\frac{m}{kg}\ =\ \frac{m^2}{s^2}\ =\ K[\kappa^A].
\ee
Consequently, we make the following choice by taking \R{Z5a} into consideration
\bey\nonumber
\mbox{\sf Setting VIII:}\hspace{5cm}
\\ \label{Z7}
\kappa^A\ \st{\td}{=}\ \frac{\mu^A}{\Theta^A}.\hspace{3cm}
\eey

According to the second term of \R{T8}  we have the following equation of physical dimensions
\bee{Z8}
[\lambda^{Ak}]\ =\ [\lambda u^{Ak}]\ =\ \frac{1}{K}\frac{m}{s}
\ee
that means, $\lambda^{Ak}$ is proportional to a velocity and at the same time perpendicular
to $u^{Ak}$ according to \R{*aT2}$_1$. Consequently, only the velocity $u^m$ of the mixture
remains for defining $\lambda^{Ak}$ in accordance with \R{*aT2}$_1$
\bey\nonumber
\mbox{\sf Setting IX:}\hspace{5cm}
\\ \label{Z10}
\lambda^{Ak}\ \st{\td}{=}\ \frac{1}{\Theta^A}u^mh^{Ak}_m.\hspace{2cm}
\eey

We know from the non-relativistic continuum theory and from \R{O2}$_1$ and \R{J6} the
following connection of the physical dimensions\footnote{angular momentum = spin density per time}
\bee{Z10+}
[k^A_l]m\ =\ \frac{N}{m^3}m\ =\ [m^{ab}]\ =\ \frac{1}{s}[s^{ab}]\ =\ \frac{N}{m^2}.
\ee
From the last term of \R{K19a} follows by taking \R{Z10+} into account
\bee{Z10+a}
[S^{Akab}]\ =\ \frac{s}{m}[s^{Aab}]\ =\ \frac{s^2}{m^3}N.
\ee
From the first term of the third row of \R{+bT2} follows by use of \R{+T2}$_1$ and \R{T2w}$_4$
\bee{Z10+b}
[S^{Ak}]\ =\ [\Lambda^A_a]\frac{m}{s}[S^{Akab}]\ =\  [\Lambda^A_{ab}][S^{Akab}]\ =\ 
 \frac{Nm}{m^3}\frac{m}{s}\frac{1}{K},
\ee
and taking \R{Z10+a} into account, we obtain
\bee{Z10+c}
[\Lambda^A_a]\ =\ \frac{1}{K}\frac{m}{s^2},\qquad [\Lambda^A_{ab}]\ =\ 
\frac{1}{K}\frac{m}{s^2}\frac{m}{s}.
\ee
In accordance with \R{*aT2}$_{2,3}$ and analogously to \R{Z10}, the relations \R{Z10+c} allow
the following
\bey\nonumber
\mbox{\sf Setting X:}\hspace{12cm}
\\ \label{Z10+d}
\Lambda^A_a\ \st{\td}{=}\ \frac{\st{\td}{u}\!{^A_a}}{\Theta^A},\qquad
\Lambda^A_{ab}\ \st{\td}{=}\ 
\frac{1}{\Theta^A}\st{\td}{u}\!{^A_{[m}}u_{n]}h^{Am}_ah^{An}_b\ =\
\frac{1}{2\Theta^A}(\st{\td}{u}\!^A_mu_n-\st{\td}{u}\!^A_nu_m)h^{Am}_ah^{An}_b.
\eey

Inserting the Lagrange multipliers into the expression of entropy density \R{T10a}, of entropy flux
density \R{T7} and of entropy supply \R{T8}, we obtain by use of \R{K19b}
\byy{Z14}
s^A &=& \frac{1}{\Theta^A}\Big(\mu^A\rh^A + \nu^A(e^A+p^A)
+u_mp^{Am}+\st{\td}{u}\!{^A_m}\frac{1}{c^2}\Xi^{Am}
+\st{\td}{u}\!{^A_{[a}}u_{b]}\frac{1}{c^2}s^{Aab}\Big),
\\ \label{Z15}
s^{Ak} &=&\frac{1}{\Theta^A}\Big(\nu^Aq^{Ak}
+u_mt^{Akm}+\mu^AJ^{Am}h^{Ak}_m
+\st{\td}{u}\!{^A_m}\Xi^{Akm}
+\st{\td}{u}\!{^A_{[a}}u_{b]}s^{Akab}\Big),
\\ \nonumber
\varphi^A &=& \frac{1}{\Theta^A}\Big\{
\mu^A{^{(ex)}}\Gamma^A+\Big(\nu^Au^A_l+u_mh^{Am}_l\Big)k^{Al}+
\\ \label{Z16}
+\Big(\st{\td}{u}{^A_{b}}u^A_a
+\st{\td}{u}\!{^A_{[m}}u_{n]}h^{Am}_ah^{An}_b\Big)\frac{1}{c^2}m^{Aab}\Big\},
\hspace{-13cm}
\eey
The entropy production density \R{Y15w} results by inserting the Lagrange multipliers
\R{Z3},\R{Z7}, \R{Z10} and  \R{Z10+d}
\bey\nonumber
\sigma^A\ =\ \hspace{-.5cm}
&&\Big(\frac{\mu^A}{\Theta^A}\Big)_{;k}J^{Am}h^{Ak}_m
+\frac{\mu^A}{\Theta^A}\Big[{^{(in)}}\!\Gamma^A+\Big(J^{Am}h^{Ak}_m\Big){_{;k}}\Big]+
\\ \nonumber
&&+\Big(\frac{\nu^A}{\Theta^A}\Big)_{;k}q^{Ak}
+\frac{\nu^A}{\Theta^A}u^A{_{l;k}}\Big(\pi^{Akl}+u^{Ak}p^{Al}\Big)+
\\ \nonumber
&&+\ \Big(\frac{1}{\Theta^A}u^mh^A_{ml}\Big)_{;k} \Big(\frac{1}{c^2}e^Au^{Ak}u^{Al}
+\frac{1}{c^2}q^{Ak}u^{Al}+t^{Akl}\Big)+
\\ \nonumber
&&+\Big(\frac{\st{\td}{u}\!{^A_b}}{\Theta^A}\Big)_{;k}\Xi^{Akb}
-\ub{\frac{\st{\td}{u}\!{^A_b}}{\Theta^A}u^A_a}\Big(s^{Akab}{_{;k}}
+\ub{ \frac{1}{c^2}\st{\td}{s}\!{^{Aab}}}\Big)+
\\ \nonumber
&&+\Big(\frac{1}{\Theta^A}\st{\td}{u}\!{^A_{[m}}u_{n]}h^{Am}_ah^{An}_b\Big)\td
\frac{1}{c^4}u^{A[a}\Xi^{Ab]}+
\\ \label{T9z}
&&+\Big(\frac{1}{\Theta^A}\st{\td}{u}\!{^A_{[m}}u_{n]}h^{Am}_ah^{An}_b\Big)_{;k}
\Big(\frac{1}{c^2}u^{A[a}\Xi^{Akb]}+s^{Akab}\Big).
\eey
The underbraced terms result in
\bee{+T9z}
\st{\td}{u}\!{^A_b}u^A_a\st{\td}{s}\!{^{Aab}}\ =\ 
\st{\td}{u}\!{^A_b}(\ub{u^A_as^{Aab}}_{=0})^\td - \st{\td}{u}\!{^A_{[b}}\st{\td}{u}\!{^A_{a]}}s^{Aab}\ =\ 0,
\ee
that means, the spin density does not appear in the entropy production.

The first four terms of the entropy production describe the four classical reasons of irreversibility:
diffusion, chemical reactions, heat conduction and internal friction with a by the momentum flux
density modified non-equilibrium viscous tensor. The fifth term of \R{T9z}\footnote{which
vanishes in equilibrium and for free 1-component systems, as we will see below} is typical for an
interacting $^A$-component as a part of the mixture because it contains the 4-velocity of the
mixture $u^m$. The same is true for the last two spin terms which vanish for 1-component systems.
In any case, all spin terms of the fields \R{Z14} to \R{T9z} related to entropy vanish with the
4-acceleration.

Up to now, a further phenomenon of irreversibility was not taken into consideration: the multi-temperature relaxation which is discussed in the next section.

\subsection{Multi-temperature relaxation and the partial temperatures\label{PT}}

Because the different components of the mixture have different partial (reciprocal) temperatures
$\lambda^A,\ A=1,2,...,Z$, a multi-temperature relaxation\footnote{do not take multi-temperature
relaxation for the heat conduction which is caused by a temperature
gradient} takes place which is an irreversible phenomenon. Consequently, multi-temperature relaxation
has to be taken into account in the entropy identity by adding a suitable zero as done in \R{+aT2}.

A heat transfer $H^{AB}$ between two components of the mixture --$A$ and $B$--
takes place by multi-temperature relaxation, if the corresponding temperatures of the components are different from each other.
Consequently, the entropy exchange
between these two components is 
\bey\nonumber
\mbox{\sf Setting XI:}\hspace{11cm}
\\ \label{A10}
G^{AB}\ :=\ H^{AB}\Big(\frac{1}{\Theta^A}-\frac{1}{\Theta^B}\Big),
\quad H^{AB} = -H^{BA},\quad \st{\td}{H}\!{^{AB}} = 0,.
\eey
Here $H^{AB}$ is an energy density and $G^{AB}$ an entropy density according to \R{Z1}
\bee{A12}
[H^{AB}]\ =\ [e^{A}],\ \quad\ [G^{AB}]\ =\ 
[e ^A]\frac{1}{K}\ =\ [s^{A}].
\ee
For the $^A$-component, this results according to \R{A10}$_2$ in
\byy{A13}
H^{A} &:=& \sum_B H^{AB},\quad \sum_{AB} H^{AB}\ =\ 0,\quad
\sum_{A}H^{A}\ =\ 0,
\\ \label{A14}
G^{A} &:=& \sum_B H^{AB}\Big(\frac{1}{\Theta^A}-\frac{1}{\Theta^B}\Big),
\\ \label{A15}
\sum_A G^{A} &=& \sum_{AB} H^{AB}
\Big(\frac{1}{\Theta^A}-\frac{1}{\Theta^B}\Big)\ \neq\ 0,\ \mbox{if}\
\Theta^A\neq\Theta^B.
\eey

The entropy exchange of the $^A$-component according to multi-temperature exchange \R{A14}$_1$
and \R{A10}$_3$ has now to be introduced into the entropy identity \R{T10y}.
Because
$G^A$ has the same physical dimension as $s^{A}$ according to \R{A12}$_2$, it fits into the
first row of \R{T10y}. Therefore we add the zero
\bee{A16}
0\ =\ -\Big[u^{Ak}G^A\Big]_{;k}
+\sum_B\Big[u^{Ak} H^{AB}\Big(\frac{1}{\Theta^A}-\frac{1}{\Theta^B}\Big)\Big]_{;k}
\ee
to \R{T10y}. Taking \R{A10}$_3$ into account and inserting
\bee{A16a}
\sum_B\Big[u^{Ak} H^{AB}\Big(\frac{1}{\Theta^A}-\frac{1}{\Theta^B}\Big)\Big]_{;k}\ 
=\ u^{Ak}{_{;k}}G^A+\sum_B H^{AB}\Big(\frac{1}{\Theta^A}-\frac{1}{\Theta^B}\Big)^\td,
\ee
into \R{A16}
\bee{A16b}
0\ =\ -\Big[u^{Ak}G^A\Big]_{;k}
+\ u^{Ak}{_{;k}}G^A+\sum_B H^{AB}\Big(\frac{1}{\Theta^A}-\frac{1}{\Theta^B}\Big)^\td,
\ee
we obtain three additional terms which
can be directly introduced into the entropy identity without defining an additional Lagrange multiplier.
According to \R{T10y}, the three terms of \R{A16b} are attached as follows
\byy{A18x}
-G^{A}&\longrightarrow& \mbox{entropy density}
\\ \label{A18xa}
u^{Ak}{_{;k}}G^A&\longrightarrow& \mbox{entropy supply},
\\ \label{A16x}
\sum_B H^{AB}\Big(\frac{1}{\Theta^A}-\frac{1}{\Theta^B}\Big)^\td
&\longrightarrow&\mbox{entropy production density},
\eey

Introducing these terms as demonstrated in sect.\ref{IC1} into the entropy identity \R{Y15v},
the entropy density \R{Z14} becomes
\bee{+Z15+}
s^A\ =\ \frac{1}{\Theta^A}\Big(\mu^A\rh^A + \nu^A(e^A+p^A)
+u_mp^{Am}+\st{\td}{u}\!{^A_m}\frac{1}{c^2}\Xi^{Am}
+\st{\td}{u}\!{^A_{[a}}u_{b]}\frac{1}{c^2}s^{Aab}\Big)+G^A,
\ee
and the state space \R{Y2} is extended by $G^A/\rh$
\bee{Y2y}
{\sf z}^A\ =\ \Big(c^A,\frac{1}{\rh},\frac{e^A}{\rh},\frac{p^{Al}}{\rh},
\frac{\Xi^{Am}}{\rh},\frac{s^{Amn}}{\rh}, \frac{G^A}{\rh}\Big),
\qquad c^A\ :=\ \frac{\rh^A}{\rh}.
\ee
and consequently the Gibbs equation
\R{Y3} by $(G^A/\rh)^\td$. The Gibbs-Duhem equation \R{Y4} is untouched by including the
multi-temperature relaxation.

According to \R{A18xa} the entropy supply \R{Z16} results in
\bey\nonumber
\varphi^A\ =\ \frac{1}{\Theta^A}\Big\{
\mu^A{^{(ex)}}\Gamma^A+\Big(\nu^Au^A_l+u_mh^{Am}_l\Big)k^{Al}+\hspace{5cm}
\\ \label{+Z16+}
+\Big(\st{\td}{u}{^A_{b}}u^A_a
+\st{\td}{u}\!{^A_{[m}}u_{n]}h^{Am}_ah^{An}_b\Big)\frac{1}{c^2}m^{Aab}\Big\}
+u^{Ak}{_{;k}}G^A,
\eey
and the entropy production density \R{T9z} becomes by \R{A16x}
\bey\nonumber
\sigma^A\ &=&
\Big(\frac{\mu^A}{\Theta^A}\Big)_{;k}J^{Am}h^{Ak}_m
+\frac{\mu^A}{\Theta^A}\Big[{^{(in)}}\!\Gamma^A+\Big(J^{Am}h^{Ak}_m\Big){_{;k}}\Big]+
\\ \nonumber
&&+\Big(\frac{\nu^A}{\Theta^A}\Big)_{;k}q^{Ak}
+\frac{\nu^A}{\Theta^A}u^A{_{l;k}}\Big(\pi^{Akl}+u^{Ak}p^{Al}\Big)+
\\ \nonumber
&&+\ \Big(\frac{1}{\Theta^A}u^mh^A_{ml}\Big)_{;k} \Big(\frac{1}{c^2}e^Au^{Ak}u^{Al}
+\frac{1}{c^2}q^{Ak}u^{Al}+t^{Akl}\Big)+
\\ \nonumber
&&+\sum_B H^{AB}\Big(\frac{1}{\Theta^A}-\frac{1}{\Theta^B}\Big)^\td+
\\ \nonumber
&&+\Big(\frac{\st{\td}{u}\!{^A_b}}{\Theta^A}\Big)_{;k}\Xi^{Akb}
-\frac{\st{\td}{u}\!{^A_b}}{\Theta^A}u^A_as^{Akab}{_{;k}}+
\\ \nonumber
&&+\Big(\frac{1}{\Theta^A}\st{\td}{u}\!{^A_{[m}}u_{n]}h^{Am}_ah^{An}_b\Big)\td
\frac{1}{c^4}u^{A[a}\Xi^{Ab]}+
\\ \label{T9zp}
&&+\Big(\frac{1}{\Theta^A}\st{\td}{u}\!{^A_{[m}}u_{n]}h^{Am}_ah^{An}_b\Big)_{;k}
\Big(\frac{1}{c^2}u^{A[a}\Xi^{Akb]}+s^{Akab}\Big).
\eey

The ten terms of the entropy production density \R{T9zp} have as already discussed after 
\R{T9z}, the following meaning:
\begin{flushleft}
$\bullet\quad$diffusion:\ $(\mu^A/\Theta^A){_{;k}}J^{Am}h^{Ak_m}$,\\
$\bullet\quad$
by diffusion modified chemical reaction:\ $(\mu^A/\Theta^A)\Big({^{(in)}}\Gamma^A
+(J^{Am}h^{Ak}_m)_{;k}\Big)$\\
$\bullet\quad$heat conduction:\ $(\nu^A/\Theta^A){_{;k}}q^{Ak}$,\\
$\bullet\quad$multi-component modified internal friction:\ $(\nu^A/\Theta^A)u^A{_{l,k}}\Big(\pi^{Akl}+u^{Ak}p^{Al}\Big)$,\\
$\bullet\quad$multi-component interaction\footnote{This term vanishes in equilibrium and for 1-component systems: 
see sect.\ref{FC}}:\ 
$(u^mh^A_{ml}/\Theta^A){_{;k}} \Big(\frac{1}{c^2}e^Au^{Ak}u^{Al}
+\frac{1}{c^2}q^{Ak}u^{Al}+t^{Akl}\Big)$,\\
$\bullet\quad$multi-temperature relaxation:\ 
$\sum_B H^{AB}\Big((1/\Theta^A)-(1/\Theta^B)\Big)^\td$\\
$\bullet\quad$four terms describing entropy production by the spin $S^{Akab}$.
\end{flushleft}

\subsection{The 4-entropy}

We need the 4-entropy of the $^A$-component for describing thermodynamics of a mixture.
Starting with \R{+T2}$_1$, \R{+Z15+} and \R{Z15}, we obtain
\bey\nonumber
S^{Ak} = \Big\{\frac{1}{\Theta^A}\Big(\mu^A\rh^A + \nu^A(e^A+p^A)
+u_mp^{Am}
+\st{\td}{u}\!{^A_m}\frac{1}{c^2}\Xi^{Am}
+\st{\td}{u}\!{^A_{[a}}u_{b]}\frac{1}{c^2}s^{Aab}\Big)+G^A\Big\}u^{Ak}+
\\ \label{E0}
+ \frac{1}{\Theta^A}\Big(\nu^Aq^{Ak}
+u_mt^{Akm}+\mu^AJ^{Am}h^{Ak}_m
+\st{\td}{u}\!{^A_m}\Xi^{Akm}
+\st{\td}{u}\!{^A_{[a}}u_{b]}s^{Akab}\Big).\hspace{.4cm}
\eey
Rearranging results in
\bey\nonumber
S^{Ak} &=&
\frac{\mu^A}{\Theta^A}\Big(N^{Ak}+J^{Am}h^{Ak}_m\Big)+
\\ \nonumber
&&+\K{A}\Big\{\nu^A\Big((e^A+p^A)u^{Ak}+q^{Ak}\Big)
+u_m\Big(u^{Ak}p^{Am}+t^{Akm}\Big)\Big\}+
\\ \label{E1}
&&+\K{A}\st{\td}{u}\!{^A_m}\Big(\frac{1}{c^2}\Xi^{Am}u^{Ak}+\Xi^{Akm}\Big)
+\K{A}\st{\td}{u}\!{^A_{[a}}u_{b]}
\Big(\frac{1}{c^2}s^{Aab}u^{Ak}+s^{Akab}\Big)+G^Au^{Ak}.\hspace{1cm}
\eey

The transition from the interacting $^A$-component to the free 1-component system is considered
in sect.\ref{FC} and that to the mixture in sect.\ref{TM}. All quantities introduced up to here are non-equilibrium ones, because we did not
consider equilibrium conditions up to now. This will be done in the next section.

\subsection{Equilibrium}
\subsubsection{Equilibrium conditions\label{EQ}}

Equilibrium is defined by {\em equilibrium conditions} which are divided into
{\em basic} and {\em supplementary} ones \C{MUBO,MUBO1}. The basic equilibrium
conditions are
given by vanishing entropy production, vanishing entropy flux density and vanishing
entropy supply\footnote{The sign $\doteq$ stands for a setting which implements an
equilibrium condition.}:
\bee{P23} 
\sigma^{A}_{eq}\ \doteq\ 0\quad\wedge\quad s^{Ak}_{eq}\ \doteq\ 0\quad\wedge\quad
\varphi^A_{eq}\ \doteq\ 0.
\ee
A first supplementary equilibrium condition is the vanishing of all diffusion flux densities.
According to \R{K8c}$_1$, we obtain 
\bee{P23a}
J^{Aeq}_k\ \doteq\ 0\quad\longrightarrow\quad
u^{Aeq}_k\ =\ f^A_{eq}u^{eq}_k\quad\longrightarrow\quad
c^2\ =\ \ f^A_{eq}u^{eq}_k u^{Ak}_{eq}.
\ee
Taking \R{K6b}$_1$ into account, \R{P23a}$_3$ results in
\bee{P23b}
(f^A_{eq})^2\ =\ 1\quad\longrightarrow\quad f^A_{eq}\ =\ \pm 1.
\ee
Consequently, we have to demand beyond \R{P23a}$_1$ the supplementary equilibrium
condition that the mass densities are additive in equilibrium. We obtain according to \R{K6b}$_2$
and \R{K9a}$_2$
\bee{P23c}
\rh_{eq}\ \doteq\ \sum_A\rh^A_{eq}\ \longrightarrow\ f^A_{eq}\ =\ 1\ \longrightarrow\
w^A_{eq}\ =\ 0.
\ee
Taking \R{P23a}$_2$ and \R{Z10} into account, \R{P23c}$_2$ yields
\bee{P23d}
u^{Aeq}_k\ =\ u^{eq}_k\ \longrightarrow\
{\lambda}{^{Ak}_{eq}}\ =\ 0,\quad g^{Am}_{eq}\ =\ 0.
\ee

Further supplementary equilibrium conditions are given by vanishing covariant
time derivatives, except that of the four-velocity:
\bee{P24}
\boxplus^\bullet_{eq}\ \doteq\ 0,\qquad\boxplus\
\neq\ u^l,
\ee
that means $\st{\td}{u}\!{^l_{eq}}$ is in general not zero in equilibrium. Consequently,
the time derivatives of all expressions which contain the 4-velocity must be calculated
separately, as we will see below.  

According to \R{P24}$_1$, we obtain
\bee{P25}
\st{\td}{\rh}\!{^A_{eq}}\ =\ 0,\qquad
\Big(\frac{\nu^A}{\Theta^A}\Big)^\td_{eq}\ =\ 0,
\ee
and the (3+1)-components of the energy-momentum tensor, \R{J2} and \R{J3},
satisfy
\bee{aP25}
\st{\td}{e}{^A_{eq}}=0,\quad \st{\td}{p}{^{Al}_{eq}}=0,
\quad \st{\td}{q}{^{Ak}_{eq}}=0,\quad \st{\td}{p}{^A_{eq}}=0,
\quad \st{\td}{\pi}{^{Akl}_{eq}}=0.
\ee

Starting with \R{K6b}$_1$, we  have
\bee{P25b}
\st{\td}{f}\!{^A_{eq}}\ =\ \frac{1}{c^2}\Big(\st{\td}{u}{^{Aeq}_m}u^m_{eq}+
{u}{^{Aeq}_m}\st{\td}{u}\!{^m_{eq}}\Big).
\ee
Taking \R{P23d}$_1$ into account, this results in
\bee{bP25}
\st{\td}{f}\!{^A_{eq}}\ =\ 0\quad\longrightarrow\quad
\st{\td}{w}\!{^A_{eq}}\ =\ 0. 
\hspace{.3cm}\ee

In equilibrium, we have according to \R{P23d}$_1$ and \R{K15}
\bee{Y16d}
h^{Am}_{leq}\ =\ h^{m}_{leq},
\ee
and according to \R{Z10} resulting in
\bee{Y16c}
\lambda^{Aeq}{_{l,k}}\ =\ 
\Big(\frac{1}{\Theta^A}u_mh^{Am}_l\Big)_{;k}^{eq}\ =\ 0. 
\ee
Despite of $\Lambda^{Aeq}_a\neq 0$ according to \R{Z10+d}$_1$ and \R{P24}$_2$,
the time derivatives of the Lagrange multipliers vanish in equilibrium,
and according to \R{Y3} and \R{Y4}, Gibbs and Gibbs--Duhem equations
are identically satisfied in equilibrium, if the "shift of the time derivative"
\bee{+Y16d}
\st{\td}{\Lambda}{^A_m}\frac{\Xi^{Am}}{\rh}\ =\
\Big(\frac{u^A_m}{\Theta^A}\frac{\Xi^{Am}}{\rh}\Big)^\td
-\Lambda{^A_m}\Big(\frac{\Xi^{Am}}{\rh}\Big)^\td\ =\ 
0-\Lambda{^A_m}\Big(\frac{\Xi^{Am}}{\rh}\Big)^\td
\ee
is taken into account.

Another supplementary equilibrium condition is the vanishing of the mass production terms in
\R{K7a}$_{3,4}$
\bee{Y15a}
{^{(ex)}}\Gamma^A_{eq}\ \doteq\ 0\ \wedge\ {^{(in)}}\Gamma^A_{eq}\ \doteq\ 0
\ \longrightarrow\  \Gamma^A_{eq}\ =\ 0
\ee
Thus, we obtain from \R{K2}$_2$, \R{P25}$_1$ and \R{Y15a}$_3$ 
\bee{Y16}
\varrho^A{_{;k}}u^{Ak}+ \varrho^Au^{Ak}{_{;k}}\ =\ \Gamma^A
\quad\longrightarrow\quad u^{Ak}_{eq}{_{;k}}\ =\ 0.
\ee

The equilibrium temperature is characterized by vanishing multi-temperature relaxation
\bee{W1a}
\Theta^A_{eq}\ \doteq\ \Theta^B_{eq}\ \doteq\ \Theta^C_{eq}\ \doteq...\ =:\ \Theta_{eq}
\longrightarrow G^A\ =\ 0,\ \wedge  A. 
\ee
Often one can find in literature \C{KUI} the case of equilibrium of multi-temperature relaxation:
although out of equilibrium, only one temperature is considered in multi-component systems. This case is realistic, if the relaxation of multi-temperature relaxation to equilibrium is
remarkably faster than that of the other non-equilibrium variables \C{MUIV}.

Taking \R{P23d}$_1$, \R{J2}$_2$ and \R{W1a} into account, the entropy density \R{+Z15+} becomes in equilibrium using the shift of the time derivative
\bee{W1}
s^A_{eq}\ =\ \frac{1}{\Theta^A_{eq}}\Big(\mu^A_{eq}\rh^A _{eq}
+\nu^A_{eq}(e^A_{eq} + p^A_{eq})\Big).
\ee
Beyond the usual expression for the entropy density in thermostatics\footnote{Below, we will see that $\nu^A_{eq}=1$}, it includes an acceleration dependent spin term.
The energy density and the pressure are here defined by
the (3+1)-decomposition \R{J1} of the energy-momentum tensor.
The chemical potential is as well as the temperature introduced as a Lagrange multiplier.

Taking \R{P23d}$_1$, \R{P23a}$_1$ and \R{W1a} into account, the entropy flux density \R{Z15} vanishes in equilibrium, resulting in
\bee{Y16a}
0\ =\ \nu^Aq^{Ak}_{eq}+\st{\td}{u}\!{^{Aeq}_m}\Xi^{Akm}_{eq}\ \longrightarrow\
q^{Ak}_{eq}\ =\ 0,
\ee
using the shift of the time derivative according to \R{+Y16d} and \R{P24}.

Finally the entropy supply \R{+Z16+} results in
\bee{W2}
0\ =\ \nu^A_{eq}u^{Aeq}_lk^{Al}_{eq}
+\st{\td}{u}\!{^{Aeq}_b}u^{Aeq}_a\frac{1}{c^2}m^{Aab}_{eq},
\ee
that means, the power exchange caused by the force density and by the angular momentum
density vanishes in equilibrium.

The entropy production \R{T9zp} has to vanish in equilibrium according to the basic
equi\-li\-brium condition \R{P23}$_1$. Taking \R{Y15a}$_2$, \R{P23a}$_1$, 
\R{Y16a}, \R{aP25}$_2$, \R{P23d}$_1$ and \R{W1a} into account and using \R{P24},
\R{T9zp} results in
\bee{Y16b}
0\ =\ \frac{\nu^A_{eq}}{\Theta^A_{eq}}u^{Aeq}{_{l;k}}\pi^{Akl}_{eq}+\Big(\frac{\st{\td}{u}\!{^A_b}}{\Theta^A}\Big)_{;k}^{eq}\Xi^{Akb}_{eq}
-\frac{\st{\td}{u}\!{^{Aeq}_b}}{\Theta_{eq}}u^{Aeq}_a\Big(s^{Akab}{_{;k}}\Big)^{eq}.
\ee
The third term of the second row of \R{T9zp} vanishes by shift of the time derivative.
In equilibrium, spin terms appear in the vanishing power exchange \R{W2} and in the spin modified
internal friction \R{Y16b}.

As demonstrated, equilibrium of an $^A$-component in the mixture is described by three basic
equilibrium conditions \R{P23} and six supplementary ones: \R{P23a}$_1$, \R{P23c}$_1$,
\R{P24}, \R{Y15a}$_{1,2}$ and \R{W1a}. Often, we can find in the literature \C{M85,EU}
equilibrium conditions
which are different from those postulated here. The reason for that is, that entropy production and
supply and the entropy flux as starting-points for the basic equilibrium conditions differ from the
expressions \R{Z14} to \R{T9z}. Such different equilibrium conditions and their derivations are
considered in the next two sections.

\subsubsection{Killing relation of the 4-temperature}

Starting with \R{+T9a3}$_1$, we now consider the following relations
\byy{B-1}
(\lambda^A{_{;k}}u^A_l+\lambda^A u^A_{l;k})\frac{1}{c^2}e^Au^{Ak}u^{Al}
&=& \st{\td}{\lambda}\!{^A}e^A,
\\ \label{B-2}
(\lambda^A{_{;k}}u^A_l+\lambda^A u^A_{l;k})u^{Ak}p^{Al}
&=& -\lambda^A u^A_l\st{\td}{p}\!{^{Al}},
\\ \label{B-3}
(\lambda^A{_{;k}}u^A_l+\lambda^A u^A_{l;k})\frac{1}{c^2}q^{Ak}u^{Al}
&=& \lambda^A{_{;k}}q^{Ak},
\\ \label{B-4}
-(\lambda^A{_{;k}}u^A_l+\lambda^A u^A_{l;k})p^Ah^{Akl}
&=& -\lambda^A p^A  u^{Ak}{_{;k}},
\\ \label{B-5}
(\lambda^A{_{;k}}u^A_l+\lambda^A u^A_{l;k})\pi^{Akl}
&=& \lambda^A u^A_{l;k}\pi^{Akl}.
\eey
Taking \R{B-2}, \R{B-3} and \R{B-5} into account,  we obtain from \R{+T9a3}$_1$
\bey\nonumber
(\lambda^A u^A_l)_{;k}
\Big(T^{Akl}-\frac{1}{c^2}e^Au^{Ak}u^{Al} +p^Ah^{Akl}\Big)\ =\hspace{4.5cm}
\\ \label{B1}
=\ -\lambda^Au^{A}_l\st{\td}{p}\!{^{Al}}
+\lambda^A{_{;k}}q^{Ak}
+\lambda^Au^A{_{l;k}}\pi^{Akl}.\hspace{1cm}
\eey
Replacing the second row of \R{T9zp} by the LHS of
\R{B1} yields the entropy production of vanishing multi-temperature relaxation and vanishing spin
by taking \R{Z3} into account
\bey\nonumber
\mbox{\sf without spin:}\qquad
\sigma^A_0 &=& 
(\lambda^A\mu^A){_{;k}}J^{Am}h^{Ak}_m
+\lambda^A\mu^A\Big({^{(in)}}\Gamma^A-(J^{Am}h^{Ak}_m)_{;k}\Big)+
\\ \nonumber
&&+(\lambda^A u^A_l)_{;k}
\Big(T^{Akl}-\frac{1}{c^2}e^Au^{Ak}u^{Al} +p^Ah^{Akl}\Big)+
\\ \label{T7gy}
&&+\ \Big(\frac{1}{\Theta^A}u^mh^A_{ml}\Big)_{;k}\Big(\frac{1}{c^2}e^Au^{Ak}u^{Al}
+\frac{1}{c^2}q^{Ak}u^{Al}+t^{Akl}\Big).
\eey

Evident is that
\bee{B1a}
(\lambda^A u^A_l)_{;k}
\Big(T^{Akl}-\frac{1}{c^2}e^Au^{Ak}u^{Al} +p^Ah^{Akl}\Big)\ =\ 0
\ee
is not a sufficient condition for equilibrium because the equilibrium conditions
\R{Y15a}$_2$, \R{P23a}$_1$  and \R{P23d}$_1$ are not necessarily satisfied and the entropy production \R{T7gy} does not vanish.
 If the energy-momentum tensor is symmetric, \R{B1a} results in
\bee{B1b}
T^{Akl}=T^{Alk}:\quad
\Big[(\lambda^A u^A_l)_{;k}+(\lambda^A u^A_k)_{,l}\Big]
\Big(T^{Akl}-\frac{1}{c^2}e^Au^{Ak}u^{Al} +p^Ah^{Akl}\Big)\ =\ 0,
\ee
an expression which as well as \R{B1a} is not sufficient for equilibrium. Consequently, the
{\em Killing relation of the 4-temperature} $\lambda^A u^A_{l}$
\bee{V8}
\Big[(\lambda^A u^A_{l})_{;k} + (\lambda^A u^A_{k})_{,l}\Big]\ =\ 0
\ee 
is also not sufficient for equilibrium\footnote{a fact
which is well-known \C{MUBO1}}. 

If equilibrium is presupposed, the equilibrium conditions \R{Y15a}$_2$,
\R{P23a}$_1$  and \R{P23d}$_1$ are satisfied, the entropy production vanishes and
\byy{V9}
\mbox{\sf without spin:}\hspace{3cm}
(\lambda^A u^A_l)_{;k}^{eq}
\Big(T^{Akl}-\frac{1}{c^2}e^Au^{Ak}u^{Al} +p^Ah^{Akl}\Big)^{eq}\ =\ 0,
\\ \label{V9a}
T^{Akl}_{eq}=T^{Alk}_{eq}:\quad
\Big[(\lambda^A u^A_l)_{;k}+(\lambda^A u^A_k)_{,l}\Big]^{eq}
\Big(T^{Akl}-\frac{1}{c^2}e^Au^{Ak}u^{Al} +p^Ah^{Akl}\Big)^{eq}\ =\ 0
\eey
are necessary conditions\footnote{but as discussed, not sufficient conditions} for equilibrium
according to \R{T7gy}, if the spin is ignored. If the spin is taken into account, \R{V9} results by
use of the fifth row of \R{T9zp} in
\bey\nonumber
(\lambda^A u^A_l)_{;k}^{eq}
\Big(T^{Akl}-\frac{1}{c^2}e^Au^{Ak}u^{Al} +p^Ah^{Akl}\Big)^{eq}\ =\hspace{6cm} 
\\ \label{V9b}
=\ -\Big(\frac{\st{\td}{u}\!{^A_b}}{\Theta^A}\Big)_{;k}^{eq}\Xi^{Akb}_{eq}
+\frac{\st{\td}{u}\!{^{Aeq}_b}}{\Theta_{eq}}u^{Aeq}_a\Big(s^{Akab}{_{;k}}\Big)^{eq}.
\hspace{1.5cm}
\eey

There are different possibilities to satisfy \R{V9} and \R{V9a} which are discussed in the next section.

\subsubsection{The gradient of the 4-temperature\label{G4T}}

The necessary condition for equilibrium ignoring spin \R{V9} can be differently satisfied generating
different types of equilibria. There are three possibilities:
\bey\nonumber
&&\hspace{-4cm}\mbox{If equilibrium exists, one of the following three conditions is valid:}  
\\ \label{V10}
(\lambda^A u^A_l)_{;k}^{eq} &=& 0\ \longrightarrow\
\lambda^{Aeq}_{;k}u^{Aeq}_l+\lambda^A_{eq} u^{Aeq}_{l,k}\ =\ 0,
\\ \label{V11}
T^{Akl}_{eq} &=& \frac{1}{c^2}e^A_{eq}u^{Ak}_{eq}u^{Al}_{eq} -p^A_{eq}h^{Akl}_{eq},
\\ \label{V12}
(\lambda^A u^A_l)^{eq}_{;k}\ \neq\ 0&\wedge&
\Big[T^{Akl}_{eq}\ \neq\ \frac{1}{c^2}e^A_{eq}u^{Ak}_{eq}u^{Al}_{eq}
-p^A_{eq}h^{Akl}_{eq}\Big],\quad\mbox{and \R{V9} is valid.}
\eey
Multiplication of \R{V10}$_2$ with $u^{Al}_{eq}$ results in
\bee{V13}
\lambda^{Aeq}_{;k}\ =\ 0\ \wedge\ u^{Aeq}_{l,k}\ =\ 0,
\ee
that means, \R{V10} represents an intensified equilibrium because additionally to the usual
equilibrium conditions mentioned in sect.\ref{EQ}, \R{V13} is valid.

If \R{V11} is valid, the equilibrium exists in a perfect material whose entropy production is zero.
If the considered material is not perfect and if the equilibrium is not intensified, \R{V12} is valid,
and the question arises, whether \R{V9} can be valid under these constraints. To answer this question, we consider \R{B-1} to \R{B-5} in equilibrium. According to the equilibrium conditions, we obtain
\byy{B-6}
(\lambda^A_{;k}u^A_l+\lambda^A u^A_{l,k})^{eq}\frac{1}{c^2}e^Au^{Ak}_{eq}u^{Al}_{eq}\
&=& 0,
\\ \label{B-7}
(\lambda^A_{;k}u^A_l+\lambda^A u^A_{l,k})^{eq}u^{Ak}_{eq}p^{Al}_{eq}\
&=& 0,
\\ \label{B-8}
(\lambda^A_{;k}u^A_l+\lambda^A u^A_{l,k})^{eq}\frac{1}{c^2}q^{Ak}_{eq}u^{Al}_{eq}\
&=& 0,
\\ \label{B-9}
-(\lambda^A_{;k}u^A_l+\lambda^A u^A_{l,k})^{eq}p^Ah^{Akl}_{eq}\
&=& 0,
\\ \label{B-10}
(\lambda^A_{;k}u^A_l+\lambda^A u^A_{l,k})^{eq}\pi^{Akl}_{eq}\
&=& 0.
\eey
Summing up \R{B-6} to \R{B-10} yields
\bee{I9}
(\lambda^A u^A_{l}){^{eq}_{;k}}T_{eq}^{Akl}\ =\ 0.
\ee
Consequently, \R{V9} is satisfied because each of the three terms vanishes for its own, thus being compatible with \R{V12}.
If an $^A$-component of a mixture is in equilibrium, two types of equilibria can occur: one in an
arbitrary material showing the usual equilibrium conditions and another one which has beyond the
the usual equilibrium conditions vanishing temperature gradient and vanishing 4-velocity gradient
according to \R{V13}.

Evident is that an 1-component system which does not interact with other components is as a
special case included in the theory of an $^A$-component in the mixture. This case is discussed in the next section.

\section{Special Case: 1-Component System\label{FC}}
\subsection{Entropy flux, -supply and -density}

An 1-component system\footnote{that is not a mixture which is a
multi-component system by definition}
can be described by setting equal all component indices of a multi-component system
\bee{O3x} 
A, B, C,...,Z\quad\longrightarrow\quad 0,
\ee
and for shortness, we omit this common index 0. Then the basic fields of an 1-component
system are according to \R{K3}
\bee{O4x}
\mbox{rest mass density and 4-velocity:}\hspace{.5cm}\{\rh,u_k\}.
\ee
The equations \R{K5} of Setting I change into identities. According to \R{K6b}$_1$, \R{K8c}$_1$,
\R{K9a}$_2$, \R{O2} and \R{K19m}$_1$, we have
\bee{O5x}
f\ =\ 1,\quad J_k\ =\ 0,\quad w\ =\ 0,\quad T^{kl}{_{;k}}\ =\ k^l\quad
S^{kab}{_{;k}}\ =\ \frac{1}{c^2}m^{ab}.
\ee
The Lagrange multipliers become acording to \R{Z3}, \R{Z7}, \R{Z10} and \R{Z10+d}
\bee{O6x}
\lambda\ =\ \frac{\nu}{\Theta},\qquad \kappa\ =\ \frac{\mu}{\Theta},\qquad\lambda^k\ =\ 0,
\qquad \Lambda_a\ =\ \frac{\st{\td}{u}_a}{\Theta}\qquad\Lambda_{ab}\ =\ 0.
\ee

The entropy density \R{Z14} and the state space \R{Y2} are as in equilibrium of the
$^A$-component \R{W1}
\bee{O7x}
s\ =\ \frac{1}{\Theta}\Big(\mu\rh +\nu(e + p) +\st{\td}{u}_m\frac{1}{c^2}\Xi^m\Big),
\qquad{\sf z}\ =\ (\rh, e, \Xi^m).
\ee

The entropy flux \R{Z15}, the entropy supply \R{Z16} and the entropy production \R{T9zp}
are by taking \R{B1} into account\footnote{There are no chemical reactions in 1-component systems.}
\byy{O8x}
s^k\ =\ \frac{1}{\Theta}\Big( \nu q^k + \st{\td}{u}_m\Xi^{km}\Big),\quad
\varphi\ =\ \frac{1}{\Theta}\Big(\mu\ {^{(ex)}}\Gamma + \nu u_lk^l 
+\st{\td}{u}_bu_a\frac{1}{c}m^{ab}\Big),
\\ \label{O8xy}
\sigma\ =\ \Big(\frac{\nu}{\Theta} u_l\Big)_{;k}\Big(T^{kl}
-\frac{1}{c^2}eu^ku^l+ph^{kl}\Big)
+\Big(\frac{\st{\td}{u}\!{_b}}{\Theta}\Big)_{;k}\Xi^{kb}
-\frac{\st{\td}{u}\!{_b}}{\Theta}u_as^{kab}{_{;k}}.
\eey
According to sect.\ref{COMI}, the (3+1)-components of the mixture change into the corresponding
quantities of the 1-component system. The necessary equilibrium conditions of an 1-component
system are equal to those of an $^A$-component in the mixture\footnote{in equilibrium: "all cats
are grey"}.

\subsection{Equilibrium and reversibility}

Vanishing entropy production out of equilibrium
\bee{O8y}
\sigma^{rev}\ =\ 0,\qquad \Big(s^k\ \neq\ 0\ \vee\ \varphi\ \neq\ 0\Big)
\ee
belongs to reversible processes and vice versa \C{MUAS}. According to \R{O8xy}, 
\bey\nonumber
\Big(\frac{\nu}{\Theta} u_l\Big)_{;k}^{rev}
\Big(T^{kl}-\frac{1}{c^2}eu^{k}u^{l} +ph^{kl}\Big)^{rev}\ =\hspace{6cm} 
\\ \label{V9bx}
=\ -\Big(\frac{\st{\td}{u}\!{_b}}{\Theta}\Big)_{;k}^{rev}\Xi^{kb}_{rev}
+\Big(\frac{\st{\td}{u}\!{_b}}{\Theta}u_a\Big)^{rev}\Big(s^{kab}{_{;k}}\Big)^{rev}.
\hspace{1.5cm}
\eey
is sufficient and necessary for vanishing entropy production in 1-component systems.  
But concerning equilibrium, \R{V9bx} is as well as \R{V9b} only necessary but not sufficient for it.
Thus, all results of sect.\ref{G4T} change into those
of an 1-component system, if the component index $^A$ is omitted, {\em eq} is replaced by {\em rev}, equilibrium is not presupposed and the generated expressions belong to reversible processes and vice versa.

Comparing \R{V9bx} with \R{V9b} and ignoring the spin, the derivative of the 4-temperature
and the Killing relation of the 4-temperature 
\bee{O23x}
\mbox{\sf without spin:}\qquad
(\lambda u_l)_{;k}^{rev}\ =\ 0,\qquad\mbox{or}\quad T^{kl} = T^{lk}\!:\ 
\Big((\lambda u_l)_{;k}+(\lambda u_k)_{,l}\Big)^{rev}\ =\ 0
\ee 
are rather conditions for reversible processes in 1-component systems because the entropy
production is enforced to be zero without existing equilibrium. 
Independently of the 4-temperature, we obtain the well-known fact \C{STE} that
according to \R{O8xy} 
all processes of perfect materials are reversible in 1-component systems without spin
\bee{O8xc}
T^{kl}_{per}\ :=\ \frac{1}{c^2}eu^ku^l-ph^{kl}\ \longrightarrow\ \sigma_{per}\ =\ 
\Big(\frac{\st{\td}{u}\!{_b}}{\Theta}\Big)_{;k}\Xi^{kb}
-\frac{\st{\td}{u}\!{_b}}{\Theta}u_as^{kab}{_{;k}}.
\ee

\section{Thermodynamics of a Mixture\label{TM}}

According to the mixture axiom in sect.\ref{MIX}, the balance equations of a mixture look like those of an 1-component system. But a mixture as a whole behaves differently from
the interacting $^A$-component in the mixture and also differently from
an 1-component system which both were discussed in sect.\ref{IC} and sect.\ref{FC}.
Because the interaction between the components is still existing in the mixture, the
diffusion fluxes and also the multi-temperature relaxation do not vanish as in 1-component systems. Because component indices $^A$ do not appear in the
description of mixtures, they are summed up in contrast to 1-component systems for which they vanish. The Settings I to III enforce the mixture axiom resulting in
\byy{Q-1}
\mbox{mass balance:}&\quad& N^{Ak}{_{;k}}\ =\ \Gamma^A\ \longrightarrow\ 
N^{k}{_{;k}}\ =\ \Gamma,
\\ \label{Q-2}
\mbox{energy balance:}&\quad& u^A_lT^{Akl}{_{;k}}ß =\ \Omega^{A}\ \longrightarrow\ 
u_l{\sf T}^{kl}{_{;k}}\ =\ \Omega,
\\ \label{Q-3}
\mbox{momentum balance:}&\quad& h^{Am}_lT^{Akl}{_{;k}}\ =\ \Omega^{Am}\
\longrightarrow\ h^{m}_l{\sf T}^{kl}{_{;k}}\ =\ \Omega^m,
\\ \label{Q-3a}
\mbox{spin balance:}&\quad&S^{Akab}{_{;k}}\ =\ \frac{1}{c^2}m^{Aab}\
\longrightarrow\ {\sf S}^{kab}{_{;k}}\ =\ \frac{1}{c^2}m^{ab.}
\eey

The Settings I to XI are concerned with the balance equations \R{Q-1} to \R{Q-3a}, with
the entropy density, the entropy flux density, the entropy supply, with the Lagrange
multipliers and the multi-temperature relaxation. Obviously, we need an additional
setting concerning the entropy of the mixture which will be formulated in the next
section.

\subsection{Additivity of 4-entropies}
\subsubsection{Entropy density and -flux}

Like the additivity of the mass flux densities \R{K5}$_1$, the energy-momentum tensors
\R{K11c}$_2$ and the spin tensors \R{K19g}
of the $^A$-components, we demand that also of the 4-entropies are additive
\bey\nonumber
\mbox{\sf Setting XII:}\hspace{6cm}
\\ \label{Q1}
{\sf S}^k\ \st{\td}{=}\ \sum_ A S^{Ak}.   \hspace{3cm}
\eey
Consequently, we obtain from \R{E1} by use of \R{L3}$_2$ and \R{K11b}
\bey\nonumber
{\sf S}^k = \sum_A\Big\{\frac{\mu^A}{\Theta^A}\Big((1-w^A)N^{Ak}+J^{Ak}\Big)
+\K{A}\Big(\nu^AQ^{Ak}+u_m\tau^{Akm}\Big)
+\frac{\nu^A}{\Theta^A}p^Au^{Ak}+\hspace{.8cm}
\\ \label{E3}
\qquad\quad
+\K{A}\st{\td}{u}\!{^A_m}\Big(\frac{1}{c^2}\Xi^{Am}u^{Ak}+\Xi^{Akm}\Big)
+\K{A}\st{\td}{u}\!{^A_{[a}}u_{b]}
\Big(\frac{1}{c^2}s^{Aab}u^{Ak}+s^{Akab}\Big)+G^{A}u^{Ak}\Big\}.\hspace{.5cm}
\eey

According to \R{+T2z}, we obtain the entropy density and the entropy flux density
of the mixture by use of \R{K9}, \R{K6b}$_1$, \R{L1} and \R{K15a1}$_1$
\bey\nonumber
{\sf S}^ku_k = c^2{\sf s} &=&
\sum_A\Big\{
\frac{\mu^A}{\Theta^A}\Big(1-w^A\Big)\rh^Ac^2f^A
+\K{A}\Big(\nu^AQ^{Ak}+u_p\tau^{Akp}\Big)u_k+,
\\ \nonumber
&&\qquad+\Big(\frac{\nu^A}{\Theta^A}p^A+G^A\Big)c^2f^A+
\K{A}\st{\td}{u}\!{^A_m}\Big(\Xi^{Am}f^A+\Xi^{Akm}u_k\Big)+
\\ \label{E4}
&&\qquad+\K{A}\st{\td}{u}\!{^A_{[a}}u_{b]}
\Big(s^{Aab}f^A+s^{Akab}u_k\Big)\Big\},
\\ \nonumber
{\sf S}^kh_k^m = {\sf s}^m &=&
\sum_A\Big\{
\frac{\mu^A}{\Theta^A}J^{Am}+
\K{A}\Big(\nu^AQ^{Ak}+u_p\tau^{Akp}\Big)h_k^m
\\ \nonumber
&&\qquad
+\Big(\frac{\nu^A}{\Theta^A}p^A+G^A\Big)c^2g^{Am}
+\K{A}\st{\td}{u}\!{^A_p}\Big(\Xi^{Ap}g^{Am}+\Xi^{Akp}h_k^m\Big)+
\\ \label{E5}
&&\qquad+\K{A}\st{\td}{u}\!{^A_{[a}}u_{b]}
\Big(\frac{1}{c^2}s^{Aab}g^{Am}+s^{Akab}h_k^m\Big)
\Big\}.\hspace{2cm}
\eey

Taking \R{K11d} into consideration, we introduce by comparing with \R{E4} and \R{E5} the
\bey\nonumber
\mbox{\sf Setting XIII:}\hspace{5cm}
\\ \label{E6}
\nu^A\ \st{\td}{=}\ f^A .   \hspace{3cm}
\eey
With this setting, the expressions of the entropy density and the entropy flux of the
mixture correspond to those which are generated by the additivity of the
energy-momentum tensors: \R{K15b} to \R{K15j}. 

Finally, we obtain the entropy and entropy flux densities of the mixture
\bey\nonumber
{\sf s} &=&
\sum_A\Big\{
\frac{\mu^A}{\Theta^A}\Big(1-w^A\Big)\rh^Af^A+
\frac{1}{c^2}\K{A}\Big(f^AQ^{Ak}+u_p\tau^{Akp}\Big)u_k+
\\ \nonumber
&&\qquad+\Big(\frac{f^A}{\Theta^A}p^A+G^A\Big)f^A+
\K{A}\st{\td}{u}\!{^A_m}\frac{1}{c^2}\Big(\Xi^{Am}f^A+\Xi^{Akm}u_k\Big)+
\\ \label{E7}
&&\qquad+\K{A}\st{\td}{u}\!{^A_{[a}}u_{b]}
\frac{1}{c^2}\Big(s^{Aab}f^A+s^{Akab}u_k\Big)
\Big\},
\\ \nonumber
{\sf s}^m &=&
\sum_A\Big\{
\frac{\mu^A}{\Theta^A}J^{Am}+
\K{A}\Big(f^AQ^{Ak}+u_p\tau^{Akp}\Big)h_k^m
\\ \nonumber
&&\qquad
+\Big(\frac{f^A}{\Theta^A}p^A+G^A\Big)c^2g^{Am}
+\K{A}\st{\td}{u}\!{^A_p}\Big(\Xi^{Ap}g^{Am}+\Xi^{Akp}h_k^m\Big)+
\\ \label{E8}
&&\qquad+\K{A}\st{\td}{u}\!{^A_{[a}}u_{b]}
\Big(\frac{1}{c^2}s^{Aab}g^{Am}+s^{Akab}h_k^m\Big)
\Big\}.\hspace{2cm}
\eey
These expressions of the entropy and entropy flux densites of the mixture are direct results of Setting XII \R{Q1}. They will be considered in sect.\ref{EFT}.

\subsubsection{Entropy supply and production density}

From \R{Q1} and \R{+T2}$_2$ follows the entropy balance equation of the mixture
\bee{Q4}
{\sf S}^k{_{;k}}\ =\ \sum_ A S^{Ak}{_{;k}}\ =\ \sum_A\Big(\sigma^A+\varphi^A\Big)\ =\ 
^\dm\sigma +^\dm\!\varphi,
\ee
satisfying  the mixture axiom. Accepting the additivity of the entropy supplies of the
$^A$-components\footnote{Supplies are caused by external influences, productions by
internal ones. That the reason why they do not mix up.}
\bey\nonumber
\mbox{\sf Setting XIV:}\hspace{6cm}
\\ \label{Q5}
^\dm\!\varphi\ \st{\td}{=}\ \sum_A\varphi^A,\hspace{3.2cm}
\eey
we obtain from \R{Q4} the additivity of the entropy productions  of the $^A$-components
\bee{Q6}
^\dm\sigma\ =\ \sum_A\sigma^A.\hspace{1.6cm}
\ee

The entropy supply of the mixture follows from \R{+Z16+}, \R{Q5} and \R{E6}
\bey\nonumber
^\dm\!\varphi\ =\ \sum_A\Big\{\frac{1}{\Theta^A}\Big[
\mu^A{^{(ex)}}\Gamma^A+\Big(f^Au^A_l+u_mh^{Am}_l\Big)k^{Al}+\hspace{5cm}
\\ \label{Q7}
+\Big(\st{\td}{u}{^A_{b}}u^A_a
+\st{\td}{u}\!{^A_{[m}}u_{n]}h^{Am}_ah^{An}_b\Big)\frac{1}{c^2}m^{Aab}\Big]
+u^{Ak}{_{;k}}G^A\Big\}.
\eey

The entropy production of the mixture follows from \R{T9zp}, \R{Q6} and \R{E6}
\bey\nonumber
^\dm\sigma\ &=& \sum_A\Big\{
\Big(\frac{\mu^A}{\Theta^A}\Big)_{;k}J^{Am}h^{Ak}_m
+\frac{\mu^A}{\Theta^A}\Big[{^{(in)}}\!\Gamma^A+\Big(J^{Am}h^{Ak}_m\Big){_{;k}}\Big]+
\\ \nonumber
&&+\Big(\frac{f^A}{\Theta^A}\Big)_{;k}q^{Ak}
+\frac{f^A}{\Theta^A}u^A{_{l;k}}\Big(\pi^{Akl}+u^{Ak}p^{Al}\Big)+
\\ \nonumber
&&+\ \Big(\frac{1}{\Theta^A}u^mh^A_{ml}\Big)_{;k} \Big(\frac{1}{c^2}e^Au^{Ak}u^{Al}
+\frac{1}{c^2}q^{Ak}u^{Al}+t^{Akl}\Big)+
\\ \nonumber
&&+\sum_B H^{AB}\Big(\frac{1}{\Theta^A}-\frac{1}{\Theta^B}\Big)^\td+
\\ \nonumber
&&+\Big(\frac{\st{\td}{u}\!{^A_b}}{\Theta^A}\Big)_{;k}\Xi^{Akb}
-\frac{\st{\td}{u}\!{^A_b}}{\Theta^A}u^A_as^{Akab}{_{;k}}+
\\ \nonumber
&&+\Big(\frac{1}{\Theta^A}\st{\td}{u}\!{^A_{[m}}u_{n]}h^{Am}_ah^{An}_b\Big)\td
\frac{1}{c^4}u^{A[a}\Xi^{Ab]}+
\\ \label{Q8}
&&+\Big(\frac{1}{\Theta^A}\st{\td}{u}\!{^A_{[m}}u_{n]}h^{Am}_ah^{An}_b\Big)_{;k}
\Big(\frac{1}{c^2}u^{A[a}\Xi^{Akb]}+s^{Akab}\Big)\Big\}.
\eey

\subsection{Partial and mixture temperatures\label{EFT}}

We now consider the positive term in the second row of the entropy density \R{E7}
\bee{M1}
\sum_A\frac{1}{\Theta^A}(f^A)^2p^A \ =\ \frac{1}{^\dm\,\Theta}\sum_A(f^A)^2p^A\ >\ 0
\ee
by which a mixture temperature $^\dm\,\Theta$ can be defined. This mixture temperature is only
a formal quantity because it is not evident that a thermometer exists which measures
$^\dm\,\Theta$: the partial temperatures are internal contact variables \C{MU14} measured by
thermometers which are selective for the temperature $\Theta^A$ of the corresponding
$^A$-component. Evident is, the measured mixture temperature is a certain mean value
of the partial temperatures of the components of the mixture \C{MUE68,DUMUE,BOGA}, but this
measured mean value may depend on the individual thermometer and may be different from
$^\dm\,\Theta$, that means, the measured temperature is not unequivocal. Different definitions of the mixture temperature can be found in literature \C{CAJO}.
But a unique mixture temperature --independent of thermometer selectivities or arbitrary definitions--
is given in the case of {\em multi-temperature relaxation equilibrium} \R{W1a}. This case is often silently presupposed in literature, if only one temperature is used in multi-component non-equilibrium systems. Only this sure case is considered in the sequel.

We now introduce the mixture quantities $\sf e$ and ${\sf q}^m$ to the entropy density $\sf s$
and the entropy flux density ${\sf s}^m$. According to \R{K15b} and \R{K15d}, we obtain 
\bee{Q2l}
\frac{1}{c^2}\sum_A\frac{1}{^\dm\,\Theta}\Big(Q^{Ak}f^A+ u_p \tau^{Akp}\Big)u_k\ =\ 
\frac{{\sf e}}{^\dm\,\Theta},\qquad
\sum_A\frac{1}{^\dm\,\Theta}\Big(Q^{Ak}f^A+ u_p \tau^{Akp}\Big)h^m_k\ =\
\frac{{\sf q}^m}{^\dm\,\Theta}.
\ee
Taking \R{M1} and \R{Q2l} into account, \R{E7} and \R{E8} result in
\bey\nonumber
{\sf s} &=&\frac{{\sf e}}{^\dm\,\Theta}+
\sum_A\Big\{
\frac{\mu^A}{\Theta^A}\Big(1-w^A\Big)\rh^Af^A+
\frac{1}{c^2}\Big(\K{A}-\frac{1}{^\dm\,\Theta}\Big)\Big(f^AQ^{Ak}+u_p\tau^{Akp}\Big)u_k+
\\ \nonumber
&&\qquad+\Big(\frac{f^A}{^\dm\,\Theta}p^A+G^A\Big)f^A+
\K{A}\st{\td}{u}\!{^A_m}\frac{1}{c^2}\Big(\Xi^{Am}f^A+\Xi^{Akm}u_k\Big)+
\\ \label{E7y}
&&\qquad+\K{A}\st{\td}{u}\!{^A_{[a}}u_{b]}
\frac{1}{c^2}\Big(s^{Aab}f^A+s^{Akab}u_k\Big)
\Big\},
\\ \nonumber
{\sf s}^m &=&\frac{{\sf q}^m}{^\dm\,\Theta}+
\sum_A\Big\{
\frac{\mu^A}{\Theta^A}J^{Am}+
\Big(\K{A}-\frac{1}{^\dm\,\Theta}\Big)\Big(f^AQ^{Ak}+u_p\tau^{Akp}\Big)h_k^m+
\\ \nonumber
&&\qquad
+\Big(\frac{f^A}{\Theta^A}p^A+G^A\Big)c^2g^{Am}
+\K{A}\st{\td}{u}\!{^A_p}\Big(\Xi^{Ap}g^{Am}+\Xi^{Akp}h_k^m\Big)+
\\ \label{E8y}
&&\qquad+\K{A}\st{\td}{u}\!{^A_{[a}}u_{b]}
\Big(\frac{1}{c^2}s^{Aab}g^{Am}+s^{Akab}h_k^m\Big)
\Big\}.\hspace{2cm}
\eey
Evident is, that partial temperatures of the components appear in all four quantities referring to
mixture entropy: entropy density \R{E7y}, entropy flux density \R{E8y}, entropy supply \R{Q7}
and entropy production density \R{Q8}. These expressions are of a more simple shape, if the
mixture is in a multi-temperature equilibrium which is considered in the next section.

\subsection{Multi-temperature relaxation equilibrium\label{MHC}}
\subsubsection{Entropy and entropy flux densities}

Presupposing  multi-temperature relaxation equilibrium \R{W1a}, the entropy density \R{E7y} and the entropy flux density \R{E8y} become
\bey\nonumber
{\sf s} &=&\frac{{\sf e}}{^\dm\,\Theta}+
\frac{1}{^\dm\,\Theta}\sum_A\Big\{
\mu^A\Big(1-w^A\Big)\rh^Af^A+(f^A)^2p^A+
\\ \label{Q3c}
&&\qquad\qquad+\st{\td}{u}\!{^A_m}\frac{1}{c^2}\Big(\Xi^{Am}f^A+\Xi^{Akm}u_k\Big)
+\st{\td}{u}\!{^A_{[a}}u_{b]}
\frac{1}{c^2}\Big(s^{Aab}f^A+s^{Akab}u_k\Big)
\Big\},
\\ \nonumber
{\sf s}^m &=&\frac{{\sf q}^m}{^\dm\,\Theta}+
\frac{1}{^\dm\,\Theta}\sum_A\Big\{
\mu^AJ^{Am}+f^Ap^Ac^2g^{Am}+
\\ \label{Q3d}
&&\qquad\qquad+\st{\td}{u}\!{^A_p}\Big(\Xi^{Ap}g^{Am}+\Xi^{Akp}h_k^m\Big)
+\st{\td}{u}\!{^A_{[a}}u_{b]}
\Big(\frac{1}{c^2}s^{Aab}g^{Am}+s^{Akab}h_k^m\Big)
\Big\}.\hspace{1cm}
\eey

The first term in the sum of \R{Q3c} can be exploited by use of the mean value theorem
according to \R{K9b} and \R{K6b}$_{1,2}$
\bee{Q3e}
\sum_A\mu^A(1-w^A)\rh^A f^A\ =\ ^\dm\mu\sum_A(1-w^A)f^A\rh^A\ 
=\ ^\dm\mu\sum_A(f^A)^3\rh^A
\ee
Consequently, the chemical potential of the mixture is
\bee{Q3f}
{^\dm}\, \mu\ :=\ \sum_A\mu^A\frac{(1-w^A)\rh^A f^A}{\sum_B(1-w^B)\rh^B f^B},
\ee
and the entropy density of the mixture \R{Q3c} yields
\bey\nonumber
{\sf s} &=& \frac{1}{^\dm\,\Theta}{\sf e}
+\frac{1}{^\dm\,\Theta}{^\dm}\mu\sum_A(f^A)^3\rh^A
+\frac{1}{^\dm\,\Theta}\sum_Ap^A(f^A)^2+
\\ \label{Q3g}
&&\qquad+\sum_A\Big\{\st{\td}{u}\!{^A_m}\frac{1}{c^2}\Big(\Xi^{Am}f^A+\Xi^{Akm}u_k\Big)
+\st{\td}{u}\!{^A_{[a}}u_{b]}
\frac{1}{c^2}\Big(s^{Aab}f^A+s^{Akab}u_k\Big)
\Big\}.
\eey

The entropy density \R{Q3g} of the mixture in multi-temperature equilibrium is similarly
constructed, but different from the expression \R{O7x} of an 1-component system: there are
the energy-, the mass-, the pressure and the spin-term.

\subsubsection{Entropy production and -supply}

The entropy supply \R{Q7} results in
\bey\nonumber
^\dm\,\varphi\ =\ \frac{1}{^\dm\,\Theta}\sum_A\Big\{
\mu^A{^{(ex)}}\Gamma^A+\Big(f^Au^A_l+u_mh^{Am}_l\Big)k^{Al}+\hspace{3cm}
\\ \label{Q9}
+\Big(\st{\td}{u}{^A_{b}}u^A_a
+\st{\td}{u}\!{^A_{[m}}u_{n]}h^{Am}_ah^{An}_b\Big)\frac{1}{c^2}m^{Aab}\Big\}.
\eey
The entropy production density \R{Q8} becomes
\bey\nonumber
^\dm\,\sigma\ &=& \sum_A\Big\{
\Big(\frac{\mu^A}{^\dm\,\Theta}\Big)_{;k}J^{Am}h^{Ak}_m
+\frac{\mu^A}{^\dm\,\Theta}\Big[{^{(in)}}\!\Gamma^A+\Big(J^{Am}h^{Ak}_m\Big){_{;k}}\Big]+
\\ \nonumber
&&+\Big(\frac{f^A}{^\dm\,\Theta}\Big)_{;k}q^{Ak}
+\frac{f^A}{^\dm\,\Theta}u^A{_{l;k}}\Big(\pi^{Akl}+u^{Ak}p^{Al}\Big)+
\\ \nonumber
&&+\ \Big(\frac{1}{^\dm\,\Theta}u^mh^A_{ml}\Big)_{;k} \Big(\frac{1}{c^2}e^Au^{Ak}u^{Al}
+\frac{1}{c^2}q^{Ak}u^{Al}+t^{Akl}\Big)+
\\ \nonumber
&&+\Big(\frac{\st{\td}{u}\!{^A_b}}{^\dm\,\Theta}\Big)_{;k}\Xi^{Akb}
-\frac{\st{\td}{u}\!{^A_b}}{^\dm\,\Theta}u^A_as^{Akab}{_{;k}}+
\\ \nonumber
&&+\Big(\frac{1}{^\dm\,\Theta}\st{\td}{u}\!{^A_{[m}}u_{n]}h^{Am}_ah^{An}_b\Big)\td
\frac{1}{c^4}u^{A[a}\Xi^{Ab]}+
\\ \label{Q10}
&&+\Big(\frac{1}{^\dm\,\Theta}\st{\td}{u}\!{^A_{[m}}u_{n]}h^{Am}_ah^{An}_b\Big)_{;k}
\Big(\frac{1}{c^2}u^{A[a}\Xi^{Akb]}+s^{Akab}\Big)\Big\}.
\eey
The meaning of each individual term of \R{Q10} was already discussed with regard to
the $^A$-component according to \R{T9zp}.

Entropy density \R{Q3g}, entropy flux density \R{Q3d}, entropy supply \R{Q9},
entropy production density \R{Q10} and chemical potential \R{Q3f} of the mixture
are represented by sums of quantities of the $^A$-components. As expected, the
(3+1)-components of the energy-momentum tensor cannot represent the mentioned
thermodynamical quantities because diffusion fluxes, chemical potentials and temperature
are not included in the energy-momentum tensor. From them, only the energy density
${\sf e}$ and the energy flux density ${\sf q}^m$ of the mixture appear in entropy and
entropy flux densities.

A temperature $^\dm\,\Theta$ of the mixture can be defined independently of multi-temperature
equilibrium. Accordding to \R{M1}, $1/^\dm\,\Theta$ is a weighted mean value of the reciprocal
partial temperatures of the components arranged with the partial pressures, a construction which
seems very special. As already mentioned in sect.\ref{EFT}, a mixture temperature is not well
defined because it depends on the component sensitivity of a thermometer.

\subsection{Total equilibrium}

Evident is that the equilibrium conditions of a mixture follow from those of the
$^A$-com\-po\-nents which we considered in sect.\ref{EQ}. Consequently, a demand of
additional equilibrium conditions for mixtures is not necessary. Presupposing the equilibrium
conditions of an $^A$-component (discussed in sect.\ref{EQ}) and multi-temperature
equilibrium, we start with the repetition
\byy{C1}
f^A_{eq}\ =\ 1,\quad w^A_{eq}\ =\ 0,\quad u^{Aeq}_k\ =\ u_k^{eq},\quad g^{Am}_{eq}\ =\ 0,
\quad ^{(ex)}\Gamma^A_{eq}\ =\ ^{(in)}\Gamma^A_{eq}\ =\ 0,
\\ \label{C2}
{\sf e}_{eq}\ =\ \sum_A e^A_{eq},\qquad {\sf q}^m_{eq}\ =\ 0,\qquad
{\sf p}^m_{eq}\ =\ \sum_A p^{Am}_{eq},\qquad {\sf t}^{jm}_{eq}\ =\ \sum_At^{Ajm}_{eq}.
\eey
Taking \R{C1} and \R{C2} into account, the entropy density \R{Q3g} of the mixture in equilibrium and the entropy flux density \R{Q3d} result in
\bee{Q3gy}
{\sf s}_{eq}\ =\ \frac{1}{^\dm\,\Theta}{\sf e}_{eq}
+\frac{1}{^\dm\,\Theta}{^\dm}\mu\rh
+\frac{1}{^\dm\,\Theta}\sum_Ap^A_{eq}+0,\qquad {\sf s}^m_{eq}\ =\ 0,
\ee
if the the shifting of the time derivative \R{+Y16d}, \R{P24}, \R{K19b} and \R{K19c} are applied.
Interesting is, that the spin terms cancel in equilibrium.

The entropy supply \R{Q9} results in equilibrium
\bee{Q9y}
^\dm\,\varphi_{eq}\ =\ \frac{1}{^\dm\,\Theta}\sum_A\Big\{
u^{Aeq}_lk^{Al}_{eq}+\st{\td}{u}{^{Aeq}_{b}}u^{Aeq}_a
\frac{1}{c^2}m^{Aab}_{eq}\Big\}\ =\ 0
\ee
according to \R{W2}.

The entropy production density \R{Q10} results in equilibrium
\bee{Q10y}
^\dm\,\sigma_{eq}\ =\ \frac{1}{^\dm\,\Theta} \sum_A\Big\{
u^{Aeq}{_{l;k}}\pi^{Akl}_{eq}-\st{\td}{u}\!{^{Aeq}_b}u^{Aeq}_a\Big(s^{Akab}{_{;k}}\Big)^{eq}\Big\}
+\sum_A\Big(\frac{\st{\td}{u}\!{^A_b}}{^\dm\,\Theta}\Big)_{;k}\Xi^{Akb}\ =\ 0,
\ee
according to \R{Y16b}.

The vanishing entropy supply of the mixture \R{Q9y} is
satisfied by the equilibrium conditions \R{Y15a}$_1$, \R{W2} and \R{Y16d}.
The vanishing entropy production of the mixture \R{Q10y} is
satisfied by the equilibrium conditions \R{Y15a}$_2$, \R{P23a}$_1$, \R{Y16a}, \R{Y16b}, \R{P24} and \R{Y16d}.

From \R{K4} follows in equilibrium analogously to \R{Y16}
\bee{O23}
u^k_{eq}{_{\ ,k}}\ =\ 0.
\ee

\subsection{(3+1)-entropy-components and spin}

If the spin is taken into consideration\footnote{that is absolutely necessary in General Relativity
Theory, see sect.\ref{GRT}}, acceleration terms appear in the entropy density and production,
\R{Q3g} and \R{Q10}, and in the entropy flux density and supply, \R{Q3d} and \R{Q9}.
The four components \R{K19b} and \R{K19c} of the spin are differently distributed over
the (3+1)-components of the entropy:

\begin{itemize}

\item the {\em entropy density} \R{+Z15+} of an $^A$-component depends on the spin density $s^{Aab}$ and on the spin density vector $\Xi^{Am}$, whereas the entropy
density of the mixture \R{E7} depends on the four spin quantities \R{K19b} and
\R{K19c}. In 1-component systems, the entropy density \R{O7x}$_1$ depends only on
the spin vector $\Xi^m$. In equilibrium, the entropy density is for all cases independent
of the spin, \R{W1} and \R{Q3gy}$_1$. 

\item the {\em entropx flux density} \R{Z15} of an $^A$-component depends on the couple stress $s^{Akab}$ and on the spin stress $\Xi^{Akm}$, whereas the entropy flux
density of the mixture \R{Q3d} depends on the four spin quantities \R{K19b} and
\R{K19c}.  In 1-component systems, the entropy flux density \R{O8x}$_1$ 
depends only on the spin vector $\Xi^m$.  In equilibrium, the entropy flux density
\R{W1} and \R{Q3gy}$_2$ vanishes and induces $q^{Ak}_{eq}=0$.

\item the {\em entropy supply} of an $^A$-component \R{+Z16+} is as well
independent of the spin as for the mixture \R{Q9} and for an 1-component system \R{O8x}$_2$. The entropy supply vanishes in equilibrium, and a connection between
the force density $k^{Al}_{eq}$ and the angular momentum density $m^{Aab}_{eq}$
is established, \R{W2} and \R{Q9y}.

\item the {\em entropy production density} \R{T9zp} and \R{Q10} does not depend on the spin density $s^{Aab}$ for an $^A$-component and for the mixture, but a
dependence upon the three other (3+1)-spin-components exists. In
1-component systems, the entropy production density \R{O8xy} depends on the
spin stress $\Xi^{kb}$ and on the couple stress $s^{kab}$. The entropy production
density vanishes in equilibrium, and a connection between the viscosity tensor
$\pi^{Akl}_{eq}$ and the spin stress and the couple stress is established, \R{Y16b}
and \R{Q10y}.

\end{itemize}

\section{Balances, Constitutive Equations and the 2$^{\bf nd}$ Law}

Up to here, a special material was not taken into account: all considered relations are
valid independently of the material which is described by constitutive equations
supplementing the balance equations. Especially, the entropy productions \R{T9zp} of
the $^A$-component and \R{Q10} of the mixture are not specified for particular
materials. There are different possibilities for introducing constitutive
equations\footnote{as an ansatz, or better by construction procedures \C{LIU,MUEH}}.
Because constitutive equations are not in the center of our
considerations, we restrict ourselves on the easiest ansatz which only serves for
elucidation of the problem: Balance equations are generally valid for all materials, that
means, they cannot be solved without choosing a special material characterized by
constitutive equations which inserted into the balance equations transform these
into a system of solvable differential equations for the wanted fields.

The entropy production of the $^A$-component \R{T9zp} is a sum of two-piece products
whose factors are so-called "fluxes" and "forces". According to \R{T9zp}, the ten
fluxes are
\bey\nonumber
{\cal Y}^A\ =\ \Big\{
J^{Am}h^{Ak}_m,\
\Big[{^{(in)}}\Gamma^A-\Big(J^{Am}h^{Ak}_m\Big)_{;k}\Big],\  
q^{Ak},\ \Big(\pi^{Akl}+u^{Ak}p^{Al}\Big),
\\ \nonumber
\Big(\frac{1}{c^2}e^Au^{Ak}u^{Al}+\frac{1}{c^2}q^{Ak}u^{Al}+t^{Akl}\Big),
H^{A{\bf B}},\Xi^{Akb},s^{Akab},
\\  \label{O24}
\frac{1}{c^4}u^{A[a}\Xi^{Ab]},
\Big(\frac{1}{c^2}u^{A[a}\Xi^{Akb]}+s^{Akab}\Big)\Big\},
\eey
and the corresponding ten forces are
\bey\nonumber
{\cal X}^A\ =\ 
\Big\{\ \Big(\frac{\mu^A}{\Theta^A}\Big)_{;k},\frac{\mu^A}{\Theta^A},\ 
\Big(\frac{f^A}{\Theta^A}\Big)_{;k},\ \frac{f^A}{\Theta^A}u^A{_{l,k}},\hspace{5cm} 
\\ \nonumber
\Big(\frac{1}{\Theta^A}u^mh^A_{ml}\Big)_{;k},\ 
\Big(\frac{1}{\Theta^A}-\frac{1}{\Theta^{\bf B}}\Big)^\td,\ 
\Big(\frac{\st{\td}{u}\!{^A_b}}{\Theta^A}\Big)_{;k},\ 
\frac{\st{\td}{u}\!{^A_b}}{\Theta^A}u^A_{a;k},\hspace{2cm}
\\ \label{O25}
\Big(\frac{1}{\Theta^A}\st{\td}{u}\!{^A_{[m}}u_{n]}h^{Am}_ah^{An}_b\Big)\td,\ 
\Big(\frac{1}{\Theta^A}\st{\td}{u}\!{^A_{[m}}u_{n]}h^{Am}_ah^{An}_b\Big)_{;k}
\Big\}.
\eey

The entropy production density \R{T9zp} of an $^A$-component can be written as
a scalar product of forces and fluxes
\bee{O26}
\sigma^A\ =\ {\cal Y}^A\cdot{\cal X}^A,
\ee
a relation which is valid independently of the material in consideration.
The material is described by the dependence of the fluxes on the forces, by the
constitutive equations 
\bee{O27}
{\cal Y}^A\ =\ {\bf F^A}({\cal X}^A)
\ee
which have to be introduced into the expression of the partial entropy
production density \R{O26} resulting in the entropy production density of the mixture
by \R{Q6}
\bee{O28}
\sigma^A\ =\ {\bf F^A}({\cal X}^A)\cdot{\cal X}^A
\qquad\longrightarrow\qquad
^\dm\,\sigma\ =\ \sum_A {\bf F^A}({\cal X}^A)\cdot{\cal X}^A\ \geq\ 0.
\ee
The inequality is caused by the Second Law which states that the entropy production of
the mixture is not negative {\em after having inserted the constitutive equations}
into the general expression \R{Q8}. Consequently, the Second Law represents a constraint
for the constitutive equations \R{O27} \C{TRIPACIMU}, and it makes no sense to take
the Second
Law into consideration before the constitutive equations are inserted. 
The entropy production of sub-systems --here the $^A$-components \R{O28}$_1$--
is not necessarily positive semi-definite. There are different methods for
exploiting the dissipation inequality \R{O28}$_2$ \C{TRIPACIMU,MUTRIPA} which are beyond this paper because special materials are here out of scope.

\section{Special Case: General Relativity Theory\label{GRT}}
\subsection{Extended Belinfante/Rosenfeld procedure}

The basic equations of General-Covariant Continuum Thermodynamics (GCCT) of a
mixture (sect.\ref{TM}) \R{Q-1} to \R{Q-3a} and \R{Q4}
\bee{C3}
N^k{_{;k}}\ =\ \Gamma, \qquad {\sf T}^{kl}{_{;k}}\ =\ k^l,\qquad 
{\sf S}^{kab}{_{;k}}\ =\ \frac{1}{c^2}m^{ab},\qquad {\sf S}^k{_{;k}}\ =\ ^{\dm}\sigma
+^{\dm}\varphi
\ee
contain covariant derivatives depending on the geometry of the space-time in which the physical
processes occur. Here, the pseudo-Riemannian space of General Relativity Theory (GRT)
is chosen as a special case.

In GRT, as a consequence of Einstein's equations
\bee{C4}
R^{ab} - \frac{1}{2}g^{ab}R\ =\ \kappa\Theta^{ab}\quad\Longrightarrow\quad
\Theta^{ab}\ =\ \Theta^{ba},\quad
\Theta^{ab}_{\ \ ;b}\ =\ 0,
\ee
the gravitation generating energy-momentum tensor $\Theta^{ab}$ has to be
symmetric and divergence-free ($R^{ab}$ is the Ricci tensor, $g^{ab}$ the metric, $R=
R^m{_m}$).
According to \R{K11c} and \R{J1}, the energy-momentum tensor
of the mixture ${\sf T}^{kl}$ may be neither symmetric nor divergence-free. The same is true for
spin divergence ${\sf S}^{kab}{_{;k}}$. Consequently, both tensors cannot serve as gravitation
genrating tensors in Einstein's equations, and the question arises: how can the balance equations
\R{C3}$_{2,3}$ be incorporated  into the general-covariant framework of GRT ?   
The answer to that question has been proved by the following extended Belinfante/Rosenfeld
procedure whose special relativistic version is well known since a long time
\C{MAT,MAT1,PAPA}. The general relativistic version is as follows
\vspace{.3cm}\newline
$\blacksquare$ {\sf Proposition}\C{MUBO14}: 
The general-covariant Belinfante/Rosenfeld procedure generates a symmetric and divergence-free
tensor 
\bee{C5}
^\dagger\Theta^{ab}\ :=\ {\sf T}^{ab}
-\frac{1}{2}\Big[{\sf S}^{kab}+{\sf S}^{abk}+{\sf S}^{bak}\Big]_{;k},
\ee
if  the GCCT balances \R{C3}$_{2,3}$ and the Mathisson-Papapetrou equations \R{C6}$_2$ and
\R{C7}$_2$
\byy{C6}  
\frac{1}{c^2}m^{ab} &\st{B}{=}& {\sf S}^{kab}{_{;k}}\ \st{{MP}}{=}\ 2{\sf T}^{[ab]},
\\ \label{C7}
k^b &\st{B}{=}& {\sf T}^{ab}{_{;a}}\ \st{{MP}}{=}\ \frac{1}{2}\Big[{\sf S}^{kab}+{\sf S}^{abk}+{\sf S}^{bak}\Big]_{;k;a}\ =\
-\frac{1}{2}R^b_{klm}{\sf S}^{klm}
\eey
(${R}^b_{klm}$ is the curvature tensor)
are valid as necessary constraints for the force density and the angular
momentum.\hfill $\blacksquare$

The Mathisson-Papapetrou equations are general-covariant including the
special-re\-la\-ti\-vistic case which is characterized by replacing the covariant derivatives by
commuting partial ones and by $R^b_{klm}\equiv 0$.

Inserting \R{C6} into \R{C5} results in
\bee{C8}
^\dagger\Theta^{ab}\ =\ {\sf T}^{(ab)} -\frac{1}{2}\Big[{\sf S}^{abc}+{\sf S}^{bac}\Big]_{;c}\
=\ {\sf T}^{(ab)}-{\sf }{\sf S}^{(ab)c}{_{;c}},
\ee
a tensor which is symmetric and divergence-free according to \R{C5} and \R{C7}$_2$.

The general-covariant Belinfante/Rosenfeld procedure transforms by use of the symmetric
spin divergence ${\sf S}^{(ab)c}{_{;c}}$ the not necessary divergence-free 
symmetric part of the energy-momentum tensor ${\sf T}^{(ab)}$ into a symmetric and 
divergence-free tensor $^\dagger\Theta^{ab}$. Or in other words: the energy-momentum
tensor ${\sf T}^{ab}$ (not necessary symmetric and divergence-free) is tranformed into the
mutant $^\dagger\Theta^{ab}$ (symmetric and divergence-free)\footnote{this name was
coined by H.-H. von Borzeszkowski}.

The decisive step for connecting GRT and GCCT is the following usually used
\bey\nonumber
\mbox{\sf Setting XV:}\hspace{3cm}
\\ \label{C9}
\Theta^{ab}\ \st{\td}{=}\ ^\dagger\Theta^{ab}:\hspace{1cm}
\eey
The mutant which is created by the Belinfante/Rosenfeld procedure is the gravitation
generating energy-momentum tensor of Einstein's equation \R{C4}. According to \R{C3}, the
mixture (and not single components) determines the geometry.

\subsection{Example: 2-component plain-ghost mixture}
\subsubsection{The plain component}

The balance equations defining the plain component (P) are according to \R{K2}, \R{O2} and
\R{K19m}$_1$
\byy{C10}
N^P_k &=& \varrho^P u^P_k,\qquad N^{Pk}{_{;k}}\ \st{B}{=}\ 0,
\\ \label{C11}
T^{[ab]}_P &=& 0,\qquad T^{(ab)}_P{_{;a}}\ =\ 0\ \st{B}{=}\ k^b_P,
\\ \label{C12}
S^{abc}_P &=& 0,\qquad
S^{kab}_P{_{;k}}\ =\ 0\ \st{B}{=}\ \frac{1}{c^2}m^{ab}_P .
\eey
By definition, the plain component is characterized by a symmetric and divergence-free
energy-momentum tensor and vanishing spin. According to \R{C6} and \R{C7},the
Mathisson-Papapetrou equations are satisfied by \R{C10} to \R{C12}, so that the plain component
of cause fits into GRT. If the plain component is regarded as an 1-component 
system\footnote{and not as a mixture component}, the gravitation
generating energy-momentum tensor is as expected according to \R{C8}$_2$ and \R{C9}
\bee{C12a}
\Theta^{ab}_P\ =\ T^{(ab)}_P.
\ee
The situation changes, if the plain component is regarded as a mixture component of a 2-component
mixture whose second component is introduced in the next section.

\subsubsection{The ghost component}

The balance equations defining the ghost component (G) are
\byy{C13}
N^G_k &=& \varrho^G u^G_k\ =\ 0,
\\ \label{C14}
T^{(ab)}_G &=& 0,\qquad T^{[ab]}_G{_{;a}}\ \st{B}{=}\ k^b_G,
\\ \label{C15}
S^{kab}_G{_{;k}} &\st{B}{=}& \frac{1}{c^2}m^{ab}_G,\qquad S^{(ab)k}_G{_{;k;a}}\ =\ 0.
\eey
By definition, the ghost component is characterized by vanishing mass density\footnote{no
"normal" material ($\varrho^G=0$), therefore the name "ghost"}. The Mathisson-Papapetrou
equations demand
\bee{C15a}
\frac{1}{c^2}m^{ab}_{G\ ;a}\ =\ 2k^b_G,
\ee
a dependence between force and angular momentum densities.
If the ghost component is regarded as an 1-component system, the gravitation
generating energy-momentum tensor is according to \R{C8}$_2$ and \R{C9}
\bee{C15b}
\Theta^{ab}_G\ =\ -S_G^{(ab)k}{_{;k}}.
\ee
Surprising is, that a "non-material" system such as the ghost component has a gravitational
effect. 

Plain and ghost components form a 2-component mixture which is discussed in the next section.

\subsubsection{The plain-ghost mixture}

According to the settings \R{K5}, \R{K11c} and \R{K19g}, we obtain for the mixture by taking
\R{C10} and \R{C13}, \R{C11}$_1$ and \R{C14}$_1$ and \R{C12}$_2$ into account
\byy{C16} 
N_k &=& N^P_k + N^G_k\ =\ \varrho^P u^P_k,
\\ \label{C17}
{\sf T}^{ab} &=& T^{(ab)}_P + T^{[ab]}_G\ \longrightarrow\ {\sf T}^{(ab)}\ =\ T^{(ab)}_P,
\quad  {\sf T}^{[ab]}\ =\ T^{[ab]}_G,
\\ \label{C18}
&&\hspace{2.4cm}\longrightarrow{\sf T}^{ab}{_{;a}}\ =\ T^{[ab]}_G{_{;a}},
\\ \label{C19}
{\sf S}^{abc} &=&  S^{abc}_P+ S^{abc}_G\ \longrightarrow\ {\sf S}^{(ab)c}{_{;c}}\ =\
S^{(ab)c}_G{_{;c}},
\\ \label{C20}
&&\hspace{2.4cm}\longrightarrow {\sf S}^{kab}{_{;k}}\ =\ S^{kab}_P{_{;k}}+ S^{kab}_G{_{;k}}\ =\ 
\frac{1}{c^2}m^{ab}_G.
\eey

For fitting the GCCT plain-ghost mixture into the GRT, the Mathisson-Papapetrou equations
\R{C6} and \R{C7} have to be satisfied for the mixture. Taking \R{C12}$_1$, \R{C18} and \R{C20}
into account, \R{C6} results in
\bee{C21}
\frac{1}{c^2}m^{ab}_G\ =\ 2T^{[ab]}_G.
\ee
According to \R{C11}$_{2}$ and \R{C18}$_2$, \R{C7} yields \R{C15a}.

The gravitation generating energy-momentum tensor of the plain-ghost mixture is according to
\R{C8}, \R{C17}$_2$ and \R{C19}$_2$ 
\bee{C23}
\Theta^{ab}\ =\ {\sf T}^{(ab)}-{\sf }{\sf S}^{(ab)c}{_{;c}}\ =\ T_P^{(ab)}-S^{(ab)c}_G{_{;c}}\
=\ \Theta_P^{ab} + \Theta_G^{ab}
\ee
which is different from \R{C12a} and \R{C15b}.

\subsection{"Dark matter" as a ghost component ?}

The previous statements allow to discuss the following (strange) situation: An observer
takes the plain-ghost mixture for an 1-component plain mixture  because the ghost
component is "dark": no additional mass density and no spin can be detected from the
point of view of the plain component. Erroneously, this observer supposes that \R{C10}
to \R{C12} are valid, but in fact, \R{C16} to \R{C20} are true. Observations of
gravitational effects yield that the gravitation generating energy-momentum
tensor $\Theta_P^{ab}$ \R{C12a} does not describe the observed gravitation because
the ghost component of the plain-ghost mixture is invisible for the observer according
to \R{C13}. This situation remembers lively the search for "dark matter" which should
correct the energy-momenrum tensor of the "visible matter". If the observer speculates
that the "dark matter" is a matter-free and spin-equipped object according to \R{C13}
to \R{C15}, the gravitation generating energy-momentum tensor $\Theta^{ab}$
describes the gravitation correctly in contrast to $\Theta^{ab}_P$.

One question arises: Does a ghost component exit in nature and what is its physical
essence ?

\section{Summary}

A multi-component system is formed by its components which are characterized by own individual
quantities, such as velocity, density, chemical potential, stress tensor, temperature, heat flux and
entropy flux densities, entropy production and supply and further items. All these individual
quantities determine those of the multi-component system which is described as a mixture.  
Individual temperatures of the components result in multi-temperature relaxation towards the corresponding
equilibrium generating a common temperature of all components and the mixture. A temperature
of the mixture in multi-temperature relaxation non-equilibrium depends on the used thermometer and
cannot be defined unequivocally.

Starting out with the rest mass densities of the components of the multi-component system,
the mass flux densities of the components are defined by introducing their different 4-velocities.
The mixture of the components is characterized by several settings. The first one is the additivity
of the component's mass flux densities to the mass flux density of the mixture. In combination with
the mixture axiom, this setting allows to define mass density and 4-velocity of the mixture and the
diffusion fluxes of the components. The non-symmetric energy-momentum tensor of one component interacting
with the mixture is introduced, and its (3+1)-split together with the component's mass and diffusion
flux densities are generating the entropy identity \C{MUBO}. The exploitation of the entropy identity requires additional settings: the entropy density, flux and supply. These settings result in physical
interpretations of entropy density, flux and supply. The entropy production follows from the entropy
identity which restricts possible arbitrariness of defining. 

By use of the entropy identity, Lagrange multipliers are introduced concerning the constraints taken into considerstion. These are temperature, chemical potential and
an additional non-equilibrium variable which characterizes the considered
component to be a part of the mixture. Beside the classical irreversible processes
--diffusion, chemical reactions, heat conduction and friction-- an additional irreversible
process --multi-temperature relaxation-- appears due to
the embedding of the considered component into the mixture. Different from the
classical case, the mass production term, the heat flux density and the viscous tensor
are modified by so-called effective quantities.

Equilibrium is defined by equilibrium conditions which are divided into
necessary and supplementary ones \C{MUBO,MUBO1,SCHBOCHMU}. The necessary equilibrium conditions are
given by vanishing entropy production, vanishing entropy flux density and vanishing
entropy supply. Supplementary equilibrium conditions are: vanishing diffusion flux densities,
vanishing component time derivatives\footnote{except that of the 4-velocity} and vanishing of the
mass production terms. Presupposing these equilibrium conditions, we obtain: all components have
the same 4-velocity, all heat flux densities are zero, the power as well as the divergence of the
4-velocity of each component vanish, and the viscous tensor is perpendicular to the velocity gradient.

The corresponding free component is defined by undistinguishable component
indices\footnote{that is not the mixture which is a multi-component system}. This 1-component
system represents the easiest classical case serving as a test, if the interacting component in the
mixture is correctly described. The vanishing of the entropy production in equilibrium is shortly
investigated: the so-called Killing relation of the vector of 4-temperature is neither a necessary nor
a sufficient condition for equilibrium. Also the statement that materials are perfect in equilibrium
cannot be confirmed.
\vspace{.7cm}\newline
{\bf Ackowledgement} Discussions with  H.-H. v. Borzeszkowski are gratefully acknowledged.

\section{Appendices}
\subsection{Rest mass densities\label{RMD}}

Consider two frames, ${\cal B}^A$ and ${\cal B}^B$. ${\cal B}^A$ is the rest frame of the
$^A$-component and ${\cal B}^B$ that of the $^B$-component. The corresponding rest mass
densities are
\bee{A1}
\mbox{rest mass/rest volume:}\qquad 
\rh^A\ :=\ \frac{m^A_0}{V^A_0},\qquad\rh^B\ :=\ \frac{m^B_0}{V^B_0}.
\ee
For comparing the rest mass densities, we have to choose the rest volumes to be equal
\bee{A2}
V^A_0\ \doteq\ V^B_0\ \doteq......=:\ V_0. 
\ee
By definition, the rest mass densities do not depend on the frame, that means, the rest mass
densities are {\em relativistic invariants} which should not be confused with the measured densities
in non-resting frames
\bee{A3}
{\cal B}^B:\quad\rh^A_B\ =\ \frac{\rh^A}{1-v^2_{AB}/c^2},\qquad\quad
{\cal B}^A:\quad\rh^B_A\ =\ \frac{\rh^B}{1-v^2_{BA}/c^2},\qquad v_{AB}\ =\ -v_{BA}.
\ee
Here $\rh^A_B$ is the density of the $^A$-component in the rest frame of the $^B$-component,
and $v_{AB}$ is the translational 3-velocity of ${\cal B}^A$ in the frame ${\cal B}^B$. These
densities are out of scope in this paper. 

We now consider \R{K6b}$_1$ in the rest frame ${\cal B}_O$ of the mixture which is defined by
$u^k_O = (0,0,0,c)$. Consequently, we obtain
\bee{A4a}
f^A_O\ =\ \frac{1}{c^2}u^A_{4O}c\ =\ 
\frac{1}{c}\frac{c}{\sqrt{1-v^2_{AO}/c^2}}.
\ee
Inserting \R{A4a} into \R{K6b}$_2$ results in the mass density of the mixture in its rest frame 
\bee{A4b}
{\cal B}_O:\quad
\rh_O\ =\ \sum_A f^A_O\rh^A_O\ =\ 
\sum_A\frac{1}{\sqrt{1-v^2_{AO}/{c^2}}}\frac{\rh^A}{1-v^2_{AO}/c^2}\ =\ 
\sum_A\frac{\rh^A}{(1-v^2_{AO}/c^2)^{3/2}}.
\ee
The same result is obtained, if \R{K5}$_3$ is written down for the rest system of the mixture.

\subsection{Example: Uniform component velocities\label{UCV}}

If there exists a common rest frame ${\cal B}^0$ for all
$^A$-components 
\bee{aK11}
u^A_k\ \doteq\ u^0_k,\quad \wedge A.
\ee
According to \R{K5}$_3$, we obtain
\bee{fK11}
\rh u_k\ =\ u^0_k\sum_ A\rh^A\quad\longrightarrow\quad
\rh c^2\ =\ u^ku^0_k\sum_ A\rh^A\ \wedge\ \rh u_ku^{0k}\ =\ c^2\sum_ A\rh^A,
\ee
and with \R{K6b}$_1$ follows
\bee{gK11}
\rh c^2\ =\ c^2f^0\sum_ A\rh^A\ \wedge\ \rh c^2f^0\ =\ c^2\sum_ A\rh^A
\quad\longrightarrow\quad(f^0)^2\ =\ 1,
\ee
resulting in
\bee{cK11}
f^0\ =\ \pm 1.
\ee
We obtain from \R{K6b}$_2$
\bee{dK11}
\varrho\ =\ f^0\sum_a\varrho^A\ =\ \pm \sum_a\varrho^A
\quad\longrightarrow\quad
f^0\ =\ +1,
\ee
and taking \R{fK11}$_1$ into account
\bee{eK11}
u_k\ =\ u^0_k.
\ee
As expected, the 4-velocity of the mixture is identical with the
uniform component velo\-ci\-ties.

\subsection{Stoichiometric equations\label{SE}}

The system of the relativistic stoichiometric equations runs as follows
\byy{A5}
\sum_A\nu_\alpha^A M^A_0\ =\ 0,\hspace{4.5cm}
\\ \nonumber
\mbox{component index: }A=1,2,...,Z,\qquad
\mbox{reaction index: }\alpha=1,2,...,\Omega. 
\eey
The stoichiometric coefficients $\nu_\alpha^A$ are scalars, and the partial rest mole mass $M^A_0$
is defined using the scalar mole number $n^A$ and the mole concentration $\zeta^A$ of the
$^A$-component  
\bee{A6}
M^A_0\ :=\ \frac{m^A_0}{n^A}\ =\ \frac{V_0}{n^A}\rh^A\ =\ \frac{\rh^A}{\zeta^A},\quad
\zeta^A\ :=\ \frac{n^A}{V_0}
\ee
according to \R{A1} and \R{A2}. The stoichiometric coefficients $\nu_\alpha^A$ are determined
by the partial rest mole masses $M^A_0,\ A=1,2,...,Z,$ before and after the $\alpha th$ reaction.

The time derivative of the mole number is determined by the reaction velocities
$\st{\td}{\xi}_\alpha$
\bee{A7}
\st{\td}{n}{^A}\  =\ \sum_\alpha\nu_\alpha^A\st{\td}{\xi}_\alpha.
\ee
Multiplication with $M^A_0$ results by use of \R{A5} in
\bee{A8}
M^A_0\st{\td}{n}{^A}\ =\ \sum_\alpha\nu_\alpha^A M^A_0\st{\td}{\xi}_\alpha
\ \longrightarrow\ \sum_A M^A_0\st{\td}{n}{^A}\ =\ 0.
\ee

Starting out with the physical dimensions
\bee{A9}
[\nu_\alpha^A]\ =\ mol,\ \ [n]\ =\ mol,\ \ [M^A_0]\ =\ \frac{kg}{mol},\ \ 
[\zeta^A]\ =\ \frac{mol}{m^3},\ \ [\st{\td}{\xi}_\alpha]\ =\ \frac{1}{s},
\ee
that of the mass production term in the first row of \R{T9z} is evidently
\bee{Z13a}
[{^{(in)}}\Gamma^A]\ =\ \frac{kg}{m^3s}.
\ee
A comparison with
\bee{Z13b}
[M^A_0\st{\td}{n}{^A}]\ =\ \frac{kg}{s}
\ee
shows that ${^{(in)}}\Gamma^A$ is the density which belongs to the mass production \R{A8}.
Because
according to \R{A2}, all rest mass densities are referred to the relativistic invariant $V_0$,
we obtain from \R{Z13b} and \R{Z13a} with \R{A8}
\bee{Z13c}
{^{(in)}}\Gamma^A\ =\ \frac{1}{V_0}\sum_\alpha\nu_\alpha^A M^A_0\st{\td}{\xi}_\alpha\
\longrightarrow\ \sum_A{^{(in)}}\Gamma^A\ =\ 0.
\ee

\end{document}